\documentclass[11pt]{article}
\pdfoutput=1

\usepackage{jheppub}
\usepackage{url,comment}
\usepackage{times}
\usepackage{latexsym}
\usepackage{graphicx, graphics, amsmath, amssymb, slashed, xcolor, bbm,bm,amsthm, array}
 \usepackage{subfigure}
 \usepackage{listings} 
 \usepackage{trimclip}

\usepackage{rotating}
\usepackage{afterpage}
 
\usepackage{multirow}

\usepackage{hyperref}	
\hypersetup{  colorlinks=true,     linkcolor=blue,     filecolor=magenta,          urlcolor=blue,}

\newcommand{\nc}{\newcommand}

\nc{\beq}{\begin{equation}}
\nc{\eeq}{\end{equation}}
\nc{\barray}{\begin{eqnarray}}
\nc{\earray}{\end{eqnarray}}
\nc{\barrayn}{\begin{eqnarray*}}
\nc{\earrayn}{\end{eqnarray*}}
\nc{\bcenter}{\begin{center}}
\nc{\ecenter}{\end{center}}
\nc{\mc}{\mathcal}
\nc{\er}[1]{(\ref{eq:#1})}
\nc{\onehalf}{\frac{1}{2}} 
\nc{\partialbar}{\bar{\partial}}
\nc{\psit}{\widetilde{\psi}}
\nc{\Tr}{\mbox{Tr}}
\nc{\hc}{\mbox{H.c.}}
\nc{\ev}{\;\mathrm{eV}}
\nc{\mev}{\;\mathrm{MeV}}
\nc{\gev}{\;\mathrm{GeV}}
\nc{\kev}{\;\mathrm{keV}}
\nc{\tev}{\;\mathrm{TeV}}
\nc{\pev}{\;\mathrm{PeV}}
\nc{\eev}{\;\mathrm{EeV}}

\def\chii0{\chi_i^0}
\def\chij0{\chi_j^0}

\newcommand{\gsim}{\lower.7ex\hbox{$\;\stackrel{\textstyle>}{\sim}\;$}}
\newcommand{\lsim}{\lower.7ex\hbox{$\;\stackrel{\textstyle<}{\sim}\;$}}
\nc{\ttbar}{t\bar t}

\newcommand{\sref}[1]{Section~\ref{s.#1}}

\newcommand{\cref}[1]{Chapter~\ref{c.#1}}

\graphicspath{{plots/}}

\title{
An Update to the Letter of Intent for MATHUSLA: Search for Long-Lived Particles at the HL-LHC
}

\author{
\includegraphics[width=4cm]{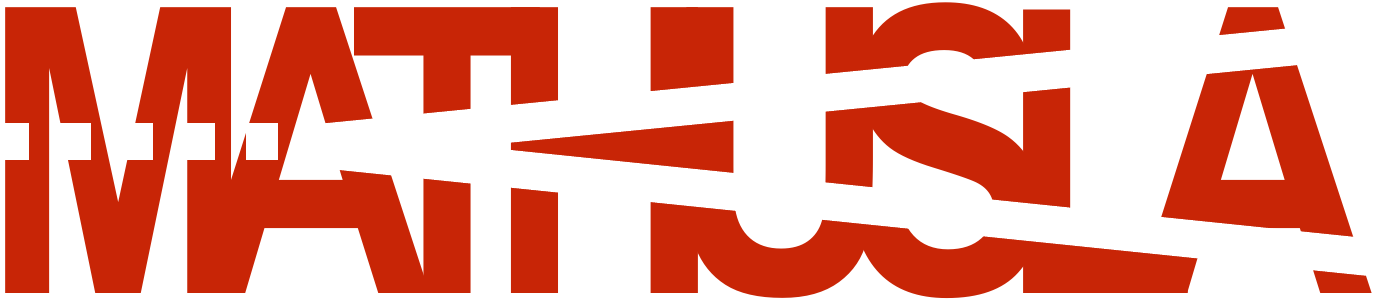}
\\
{\normalfont 
\href{https:mathusla-experiment.web.cern.ch}{\texttt{mathusla-experiment.web.cern.ch}}
\\
\vspace{5mm}}}

\author[1]{\hspace*{-1mm}Cristiano Alpigiani,}
\author[2]{Juan Carlos   Arteaga-Vel\'azquez,}
\author[3]{Austin   Ball,}
\author[4]{Liron   Barak,}
\author[5]{Jared   Barron,}
\author[6]{Brian   Batell,}
\author[7]{James   Beacham,}
\author[4]{Yan   Benhammo,}
\author[8]{Karen Salom\'e   Caballero-Mora,}
\author[9]{Paolo   Camarri,}
\author[9]{Roberto   Cardarelli,}
\author[10]{John Paul   Chou,}
\author[5]{Wentao   Cui,}
\author[5]{David   Curtin,}
\author[5]{Miriam   Diamond,}
\author[11, 12]{Keith R.    Dienes,}
\author[3]{Liam Andrew   Dougherty,}
\author[9]{Giuseppe   Di Sciascio,}
\author[13]{Marco   Drewes,}
\author[4]{Erez   Etzion,}
\author[14]{Rouven   Essig,}
\author[15]{Jared   Evans,}
\author[16]{Arturo   Fern\'andez T\'ellez,}
\author[17]{Oliver   Fischer,}
\author[18]{Jim   Freeman,}
\author[3]{Jonathan   Gall,}
\author[10]{Ali   Garabaglu,}
\author[19]{Stefano   Giagu,}
\author[10]{Stephen Elliott   Greenberg,}
\author[20]{Bhawna   Gomber,}
\author[3]{Roberto   Guida,}
\author[21]{Andy   Haas,}
\author[22]{Yuekun   Heng,}
\author[1]{Shih-Chieh   Hsu,}
\author[23]{Giuseppe   Iaselli,}
\author[11]{Ken   Johns,}
\author[1]{Audrey   Kvam,}
\author[24]{Dragoslav   Lazic,}
\author[25]{Liang   Li,}
\author[9]{Barbara   Liberti,}
\author[12]{Zhen   Liu,}
\author[1]{Henry   Lubatti,}
\author[5]{Lillian   Luo,}
\author[26]{Giovanni   Marsella,}
\author[16]{Mario Iv\'an   Mart\'inez Hern\'andez,}
\author[3]{Matthew   McCullough,}
\author[27]{David   McKeen,}
\author[14]{Patrick   Meade,}
\author[4]{Gilad   Mizrachi,}
\author[8]{O.G.   Morales-Olivares,}
\author[27]{David   Morrissey,}
\author[4]{Meny   Raviv Moshe,}
\author[19]{Antonio   Policicchio,}
\author[1]{Mason   Proffitt,}
\author[28]{Dennis Cazar   Ramirez,}
\author[29]{Matthew   Reece,}
\author[30]{Steven H.   Robertson,}
\author[16]{Mario   Rodr\'iguez-Cahuantzi,}
\author[3]{Albert   de Roeck,}
\author[31]{Amber   Roepe,}
\author[1]{Joe   Rothberg,}
\author[32]{James John   Russell,}
\author[30]{Heather   Russell,}
\author[9]{Rinaldo   Santonico,}
\author[33]{Marco   Schioppa,}
\author[34]{Jessie   Shelton,}
\author[35]{Brian   Shuve,}
\author[4]{Yiftah   Silver,}
\author[9]{Luigi   Di Stante,}
\author[36]{Daniel   Stolarski,}
\author[31]{Mike   Strauss,}
\author[37]{David   Strom,}
\author[31]{John   Stupak,}
\author[38]{Martin A.   Subieta Vasquez,}
\author[39]{Sanjay Kumar   Swain,}
\author[16]{Guillermo   Tejeda Mu\~noz,}
\author[10]{Steffie Ann   Thayil,}
\author[40]{Brooks   Thomas,}
\author[41]{Yuhsin   Tsai,}
\author[42]{Emma   Torro,}
\author[1]{Gordon   Watts,}
\author[43]{Charles   Young,}
\author[44]{Jose   Zurita}

\affiliation[1]{University of Washington, Seattle}
\affiliation[2]{Universidad Michoacana de San Nicol\'as de Hidalgo, Mexico (UMSNH)}
\affiliation[3]{CERN}
\affiliation[4]{Tel Aviv University}
\affiliation[5]{University of Toronto}
\affiliation[6]{University of Pittsburgh}
\affiliation[7]{Ohio State University}
\affiliation[8]{Universidad Aut\'onoma de Chiapas, Mexico (UNACH)}
\affiliation[9]{Sezione di Roma Tor Vergata, Roma, Italy}
\affiliation[10]{Rutgers, the State University of New Jersey}
\affiliation[11]{University of Arizona}
\affiliation[12]{University of Maryland}
\affiliation[13]{Universit\'{e} catholique de Louvain}
\affiliation[14]{YITP Stony Brook}
\affiliation[15]{University of Cincinnati}
\affiliation[16]{Benem\'erita Universidad Aut\'onoma de Puebla, Mexico (BUAP)}
\affiliation[17]{Karlsruhe Institute of Technology}
\affiliation[18]{Fermi National Accelerator Laboratory (FNAL)}
\affiliation[19]{Universit\`{a} degli Studi di Roma La Sapienza, Roma, Italy}
\affiliation[20]{Hyderabad University}
\affiliation[21]{New York University}
\affiliation[22]{Institute of High Energy Physics, Beijing}
\affiliation[23]{Politecnico di Bari, Italy}
\affiliation[24]{Boston University}
\affiliation[25]{Shanghai Jiao Tong University}
\affiliation[26]{Universit\`{a} del Salento, Lecce, Italy}
\affiliation[27]{TRIUMF}
\affiliation[28]{Universidad San Francisco de Quito (USFQ)}
\affiliation[29]{Harvard University}
\affiliation[30]{McGill University}
\affiliation[31]{University of Oklahoma}
\affiliation[32]{SLAC}
\affiliation[33]{INFN and University of Calabria}
\affiliation[34]{University of Illinois Urbana-Champaign}
\affiliation[35]{Harvey Mudd College}
\affiliation[36]{Carleton Unversity}
\affiliation[37]{University of Oregon}
\affiliation[38]{Instituto de Investigaciones F\'isicas (IIF), Observatorio de F\'isica C\'osmica de \^a Chacaltaya\^a, Universidad Mayor de San Andr\'es (UMSA)}
\affiliation[39]{National Institute of Science Education and Research, HBNI, Bhubaneswar}
\affiliation[40]{Lafayette College}
\affiliation[41]{University of Notre Dame}
\affiliation[42]{Instituto de F\'isica Corpuscular (CSIC-UV), Valencia, Spain}
\affiliation[43]{SLAC National Accelerator Laboratory}
\affiliation[44]{Karlsruhe Institute of Technology Institute for Theoretical Physics}

\emailAdd{mathusla.experiment@cern.ch}

\abstract{
We report on recent progress in the design of the proposed MATHUSLA Long Lived Particle (LLP) detector for the HL-LHC, updating the information in the original Letter of Intent (LoI), see CDS:LHCC-I-031, arXiv:1811.00927. A suitable site has been identified at LHC Point 5 that is closer to the CMS Interaction Point (IP) than assumed in the LoI. The decay volume has been increased from 20 m to 25 m in height. Engineering studies have been made in order to locate much of the decay volume below ground, bringing the detector even closer to the IP. With these changes, a 100 m x 100 m detector has the same physics reach for large c$\tau$ as the 200 m x 200 m detector described in the LoI and other studies. The performance for small c$\tau$ is improved because of the proximity to the IP. Detector technology has also evolved while retaining the strip-like sensor geometry in Resistive Plate Chambers (RPC) described in the LoI. The present design uses extruded scintillator bars read out using wavelength shifting fibers and silicon photomultipliers (SiPM). Operations will be simpler and more robust with much lower operating voltages and without the use of greenhouse gases. Manufacturing is  straightforward and should result in cost savings. Understanding of backgrounds has also significantly advanced, thanks to new simulation studies and measurements taken at the MATHUSLA test stand operating above ATLAS in 2018. We discuss next steps for the MATHUSLA collaboration, and identify areas where new members can make particularly important contributions.
}

\begin{document}

\begin{flushright}
%\phantom{\small{.}}
CERN-LHCC-2020-014\\
LHCC-I-031-ADD-1
\end{flushright}

\maketitle

%%%%%%%%%%%%%%%%%
\section{Introduction}
\label{s.introduction}
%%%%%%%%%%%%%%%%%

MATHUSLA (Massive Timing Hodoscope for Ultra-Stable neutraL pArticles)~\cite{Chou:2016lxi, Alpigiani:2018fgd} is a proposed large-scale dedicated Long-Lived Particle (LLP) detector to be situated at CERN near one of the main detectors. 
It will be able to reconstruct the decay of neutral LLPs, produced in HL-LHC collisions, as displaced vertices (DV) in a near-zero-background environment. 
The physics case for LLP searches at the HL-LHC in general and MATHUSLA in particular was explored in detail in~\cite{Curtin:2018mvb}.
MATHUSLA would be able to extend the sensitivity in long lifetime and LLP cross section by several orders of magnitude compared to the main detectors alone, depending on the production and decay mode.
In particular, it would allow searches for LLPs with lifetimes near the upper Big Bang Nucleosynthesis bound set by cosmology, and also play a vital role in the search for Dark Matter (DM).

We report several updates on the design on the MATHUSLA detector in \sref{mathuslacms}.
Site-specific engineering studies have been carried out by CERN engineers to identify a suitable location for the MATHUSLA detector on CERN-owned land adjacent to CMS. 
This informs the updated ``MATHUSLA @ CMS'' geometry, with a 100~m $\times$ 100~m area and a 25~m high decay volume that is excavated 20m below grade. 
This is only $\frac{1}{4}$ the area of the earlier MATHUUSLA200 benchmark~\cite{Chou:2016lxi, Ariga:2018zuc, Curtin:2018mvb}, a major factor in reducing costs. 
The updated location is significantly closer to the collision point than earlier benchmarks, resulting in near-identical LLP sensitivity compared to MATHUSLA200.
We also identify extruded scintillators as the leading technology choice for MATHUSLA's tracker system and are currently conducting detailed design studies. 

In \sref{sensitivity} we present new LLP sensitivity estimates for the updated MATHUSLA@CMS geometry, which confirm that the new geometry has near-identical sensitivity to MATHUSLA200. Importantly, this means that the myriad of projections in the physics case white paper~\cite{Curtin:2018mvb} can be applied almost verbatim to the updated design. 
We also draw upon recent studies from the literature to emphasize the vital role MATHUSLA plays in the hunt for Dark Matter (DM), especially in scenarios where the DM abundance is controlled by the properties of an LLP.
Finally, we summarize a recent study~\cite{Barron:2020kfo} 
demonstrating that  
MATHUSLA can characterize any discovered LLP in great detail, provided it is integrated into the CMS L1 trigger system to allow a combined analysis with main detector information.

There have been significant advances in our understanding of backgrounds to LLP searches at MATHUSLA, thanks to several detailed simulation studies as well as measurements conducted by the MATHUSLA test stand~\cite{Alidra:2020thg} in 2018. 
As a result, we can confirm earlier estimates that downward traveling cosmic rays, muons from the LHC and atmospheric neutrinos can be vetoed and are unlikely to constitute a background to LLP searches at MATHUSLA. 
On the other hand, these new results also focus attention on very rare backgrounds, namely the production of long-lived pions, muons and neutral kaons due to cosmic ray inelastic back-scattering in the detector floor. 
Several veto strategies are available, but reliable understanding of these ultra rare events necessitates careful study, which are currently in progress. 

We comment briefly on a possible upgrade to MATHUSLA in order to study cosmic rays in \sref{cosmicrays}, and then sum up the current status, as well as next steps for the collaboration, in \sref{nextsteps}. It is our aim to complete a technical design report and a robust cost estimate by early 2021, and there are ample opportunities for new collaboration members to contribute, especially in the areas of hardware design, geometry optimization, simulation studies, and cosmic ray physics.

%%%%%%%%%%%%%%%%%
\section{MATHUSLA at CMS}
\label{s.mathuslacms}
%%%%%%%%%%%%%%%%%

We now summarize recent studies in choosing a location for MATHUSLA near CMS, the updated MATHUSLA @ CMS geometry and modular design, and details of the tracking system based on extruded plastic scintillators.

\subsection{Location}

A site has been identified a LHC Point 5 close to the CMS Interaction Point (IP) that allows for a 100 m x 100 m detector footprint to be located on the CERN site at P5. The detector building and its relation to the CMS IP is shown in Figure \ref{fig:layout_P5}; the dashed red line indicates the CERN non-fenced domain boundry. The near side of this site is approximately 70 m from the CMS IP, considerably closer than the 100 m assumed in the Letter of Intent. 

\begin{figure}
\begin{center}
\includegraphics[width=0.7\textwidth]{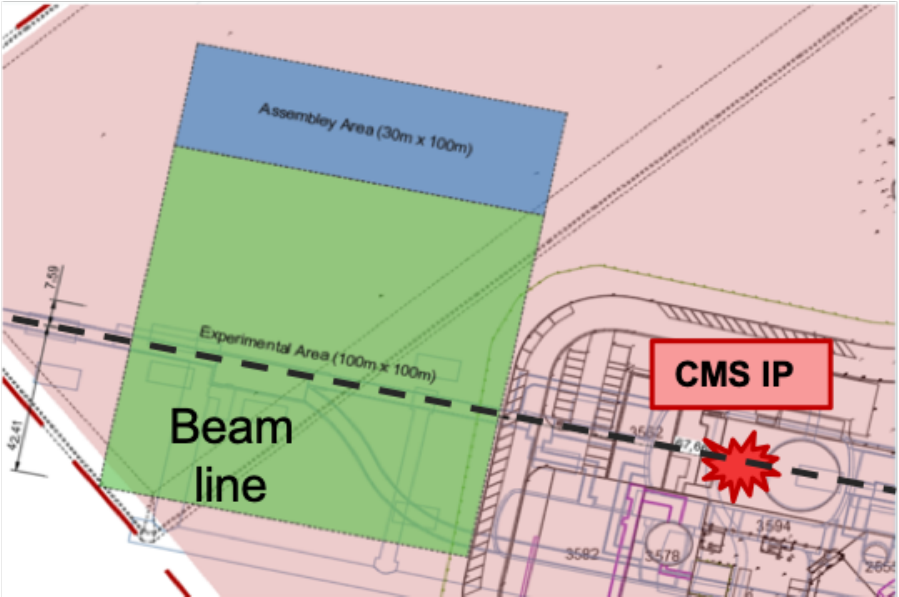}
\end{center}
\caption{Location of proposed MATHUSLA detector at the CMS site.}
\label{fig:layout_P5}
\end{figure}

\begin{figure}
\begin{center}
\includegraphics[width=1.0\textwidth]{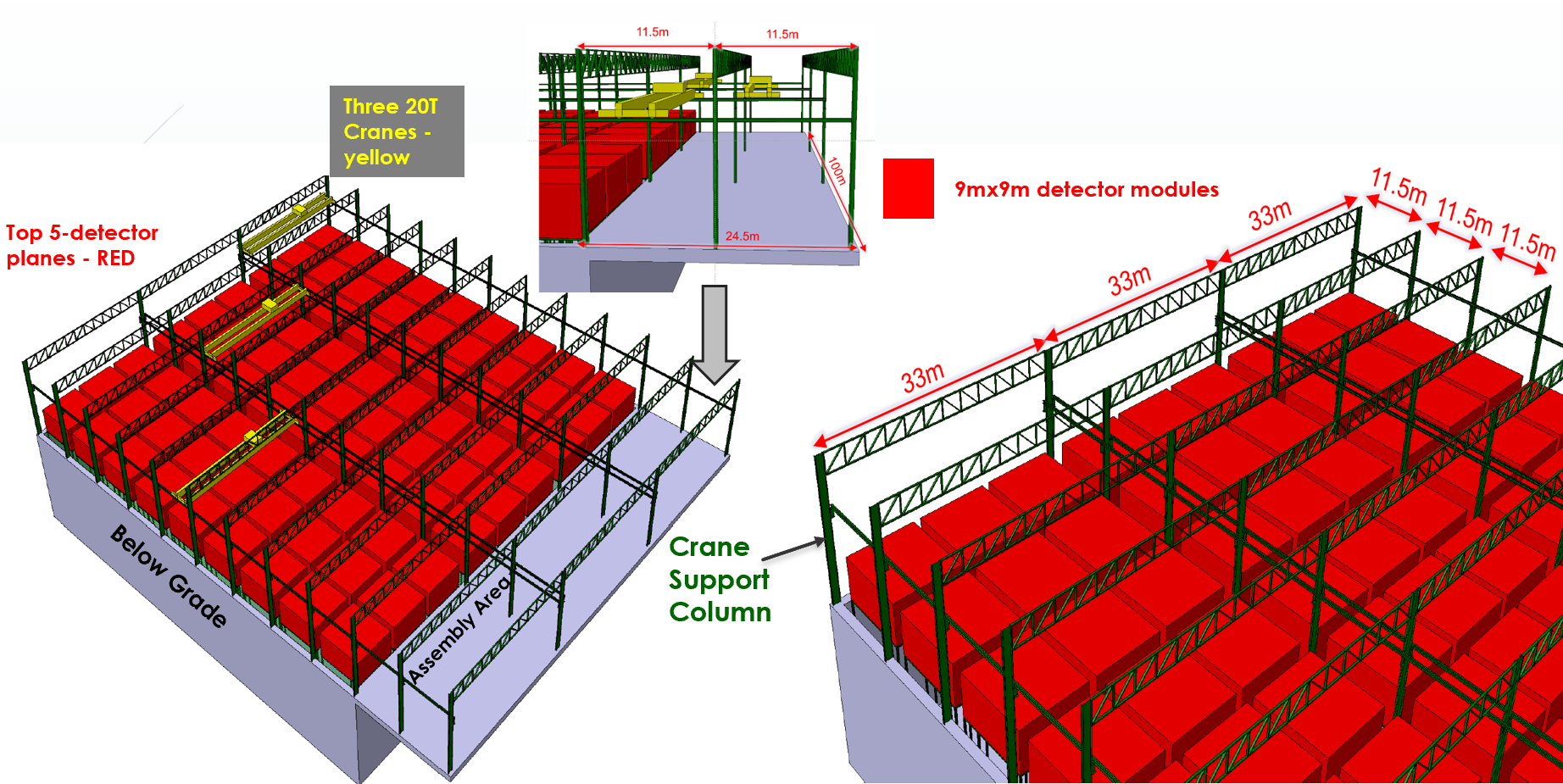}
\end{center}
\caption{Engineering details of the partially excavated structure that would house MATHUSLA.}
\label{fig:BLDG-details}
\end{figure}

Site-specific engineering studies  have been carried out by CERN engineers to locate 20 m of decay volume below grade, which together with 5 m of decay volume above ground level gives a total decay height of 25 m. This is an increase of 5 m over the design described in the LoI~\cite{Alpigiani:2018fgd}. Figure~\ref{fig:BLDG-details}\ shows the excavated volume with retaining walls and other structures. There are two rows of support that divide the volume into three regions, each of which is served by a bridge crane with 33 m span. The other columns support the detectors. At the end of the excavated volume is a surface assembly area measuring 100 m x 30 m. 

Putting the detector longitudinally closer to the IP and placing most of the decay volume below ground have increased the solid angle coverage significantly. Thus a smaller detector with 100 m x 100 m footprint in this configuration has the same physics reach as the 200 m x 200 m detector described in the LoI. We demonstrate this in \sref{sensitivity}.

\subsection{Modular detector design}

\begin{figure}
\begin{center}
\includegraphics[width=0.7\textwidth]{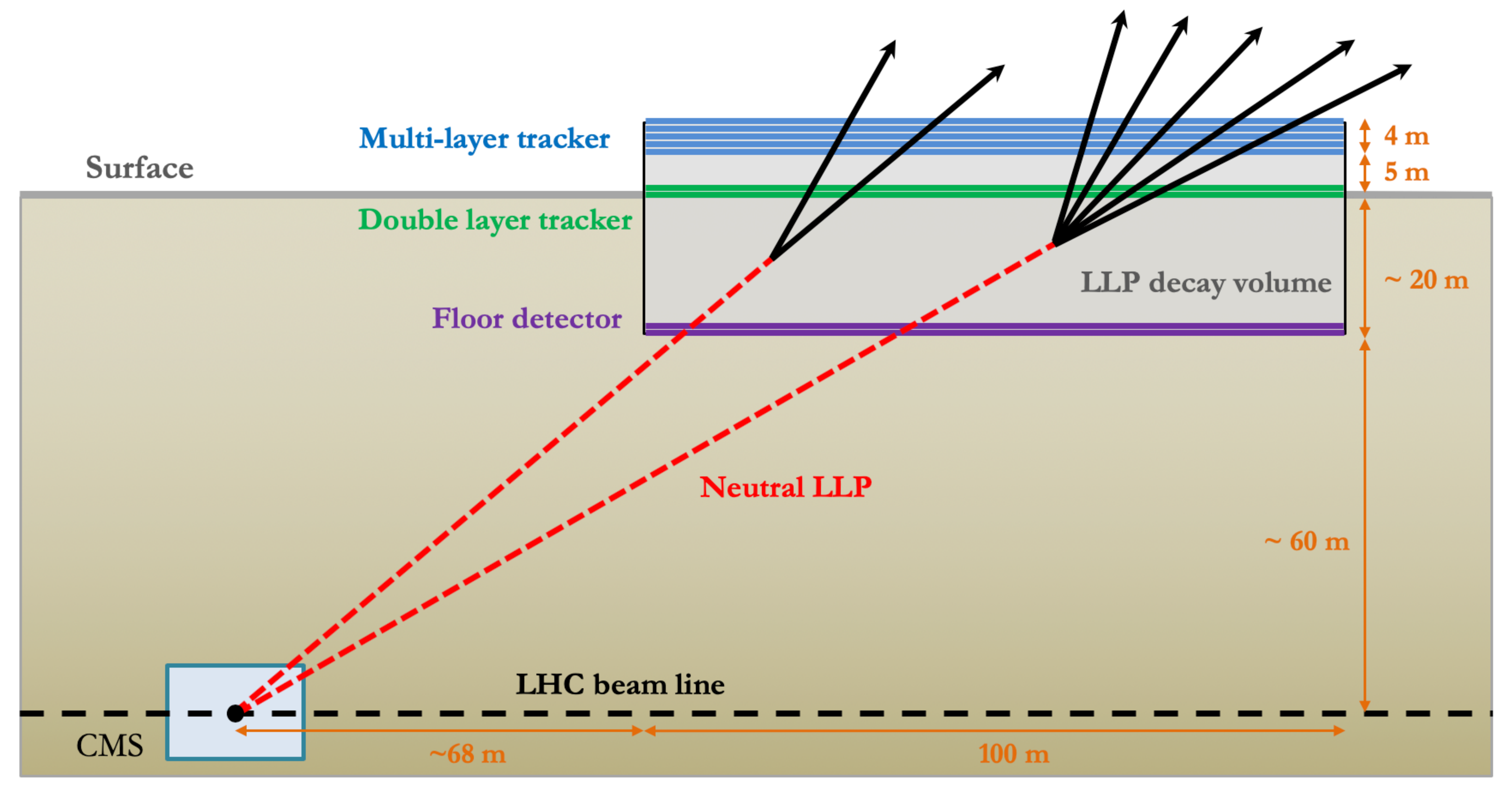}
\end{center}
\caption{MATHUSLA@CMS geometry relative to the CMS collision point.}
\label{fig:layout_v4}
\end{figure}

\begin{figure}
\begin{center}
\includegraphics[width=0.48\textwidth]{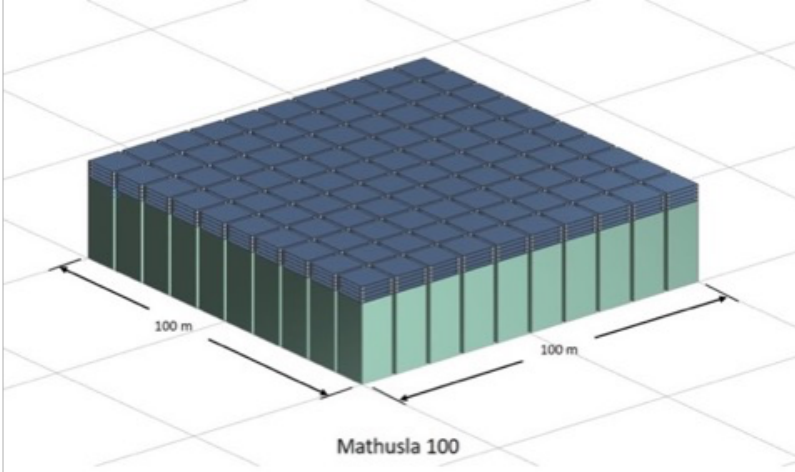}
\includegraphics[width=0.48\textwidth]{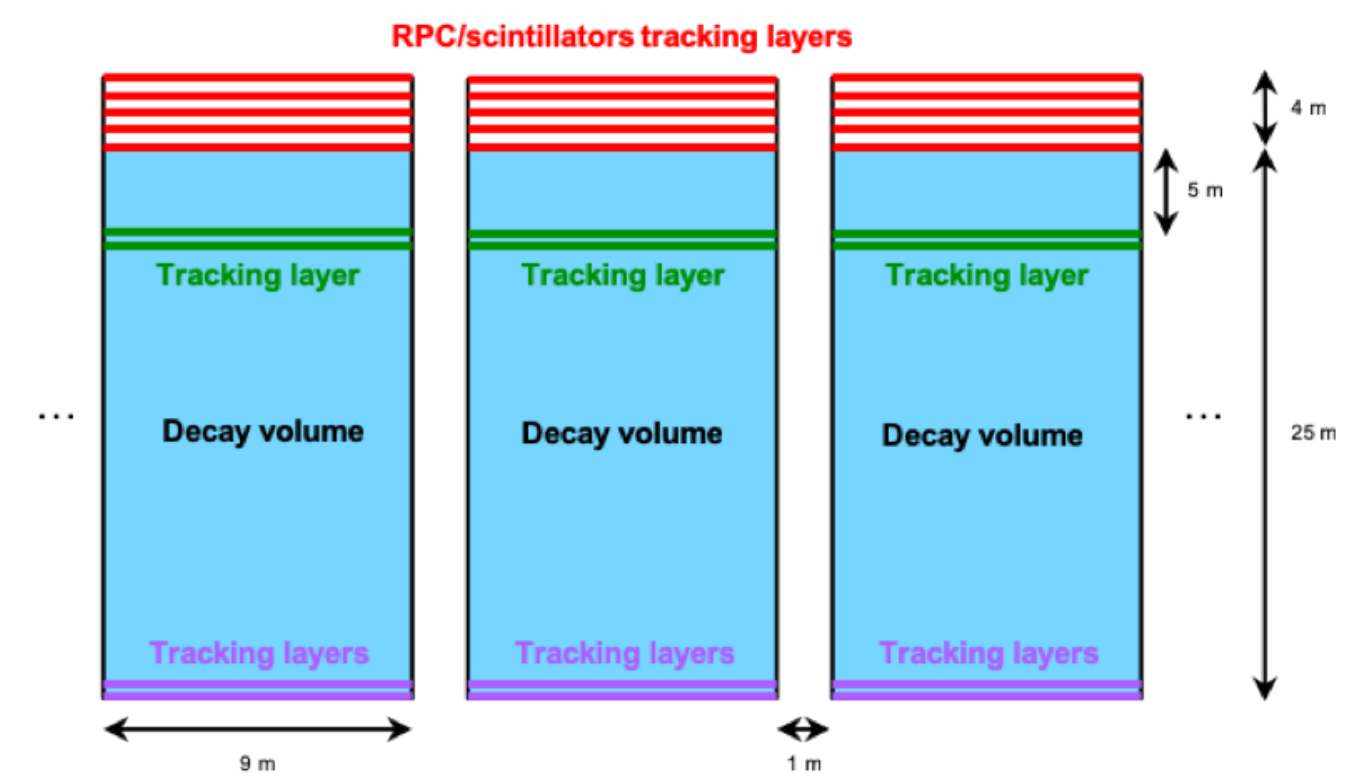}
\end{center}
\caption{Schematic view of the MATHUSLA detector modular concept: left the $100$ ($9 \,\textrm{m} \times 9 \,\textrm{m}$) units; right the detector planes in each module}
\label{fig:geometry}
\end{figure}

Further simulations studies and information from the test stand we operated at P1~\cite{Alidra:2020thg} have informed our revised MATHUSLA layout at CMS shown in Figure \ref{fig:layout_v4}. The current detector concept shown in Figure \ref{fig:geometry}\ is a \mbox{100 m$^{2}$} detector consisting of 100 $9 \,\textrm{m} \times 9 \,\textrm{m}$ units. Each detector unit comprises 9-layers of scintillating-detector planes that provide position and timing coordinates of charged particles resulting from the decay of long-lives particles in the \mbox{MATHUSLA} detector decay volume. There are five detector planes, separated by $1$ m at the top, two additional planes, also separated by $1$ m located \mbox{5 m} below (to enhance the particle position measurement precision close to the floor) and two additional planes at the floor for rejecting charged particles from the LHC and cosmic muon backscattering.  The total height of $\sim 40$ m includes a $\sim 25$ m LLP decay volume, 21 m of which would be excavated, and 12 m above the surface that hosts the tracker and the crane system used for assembly and maintenance.

The layout of the the building that houses the 100 m $\times$ 100 m experimental area and the adjacent 30 m $\times$ 100 m adjacent area for the detector assembly is shown in Figure \ref{fig:layout_P5}.  The structure, which is located on the surface near the CMS IP fits well on CERN owned land.  Having a large part of the decay volume underground brings it closer to the IP, which increases the solid angle in the acceptance for LLPs generated in the collisions. To adjust to the available land, this proposal has a 7.5 m offset to the centre of the beams. The site allows for the detector to be as close as 68 m from the IP, which is show in red in Figure~\ref{fig:geometry}.\footnote{This small offset of the detector center line from the beam is neglected in our simulation studies, since its precise parameters may change and it is unlikely to have a significant effect on LLP sensitivity.}

\iffalse
\begin{figure}[hbtp!]
\begin{center}
\includegraphics[width=0.7\textwidth]{LayoutOnP5.pdf}
\end{center}
\caption{MATHUSLA detector at CMS site}
\label{fig:layout_P5}
\end{figure}
\fi

\subsection{Scintillator Detector Planes}
\label{Scintillator.plane}

Each of the $9 \,\textrm{m} \times 9 \,\textrm{m}$ detector planes consists of an assembly of extruded scintillating bars whose length, width and thickness is  
$4.55$ m, $4.5$ cm, $2$ cm, respectively. Each bar is extruded with a hole at the centre into which  
a wave-length shifting (WLS) fibre is inserted and connected to an SiPM.  To facilitate installation the scintillating bars are assembled into $4$ sub-units that comprise $101$ bars resulting in 
$4.5 \,\textrm{m} \times 4.545 \,\textrm{m}$ sub-units that allow for overlapping the $4$ sub-units by $5$ cm longitudinally and $4.5$ cm transversely in order to avoid gaps in coverage. In this arrangement the $4$ sub-units provide hermetic coverage over a $9 \,\textrm{m} \times 9 \,\textrm{m}$ area.

%%%%%%%%%%%%%%%%%%%%%%%%%%%%%%%%%%
%%%%%%%%%%%%%%%%%%%%%%%%%%%%%%%%%%
%%%%%%%%%%%%%%%%%%%%%%%%%%%%%%%%%%
%%%%%%%%%%%%%%%%%%%%%%%%%%%%%%%%%%
%%%%%%%%%%%%%%%%%
\section{New Physics Reach of LLP Searches}
\label{s.sensitivity}
%%%%%%%%%%%%%%%%%

We now present updated (and slightly improved) LLP sensitivity estimates for the new MATHUSLA@CMS 
benchmark geometry presented above, and demonstrate that the reach is essentially identical to the old MATHUSLA200 benchmark geometry from the original LOI~\cite{Alpigiani:2018fgd}. 
We also emphasize that LLP searches are instrumental in the hunt for Dark Matter and are often the only way of observing the DM directly, as demonstrated by several recent studies~\cite{No:2019gvl,DAgnolo:2018wcn,Aielli:2019ivi,Berlin:2018jbm}.
We also summarize recent work~\cite{Barron:2020kfo}, which shows that analysis of MATHUSLA and CMS data together can characterize the LLP in great detail, including determining the production mode, decay mode, and (if applicable) parent particle mass. This serves as powerful motivation to integrate MATHUSLA with CMS so it can supply a L1 trigger signal, ensuring that the necessary information in the main detector is recorded.

%%%%%%%%%%%%%%%%%%%%%%%%%%%%%%%%%%
%%%%%%%%%%%%%%%%%%%%%%%%%%%%%%%%%%
%%%%%%%%%%%%%%%%%%%%%%%%%%%%%%%%%%
\subsection{LLP Benchmark Models}
\label{s.LLPbenchmark}

\begin{figure}
    \centering
    \includegraphics[width=0.6\textwidth]{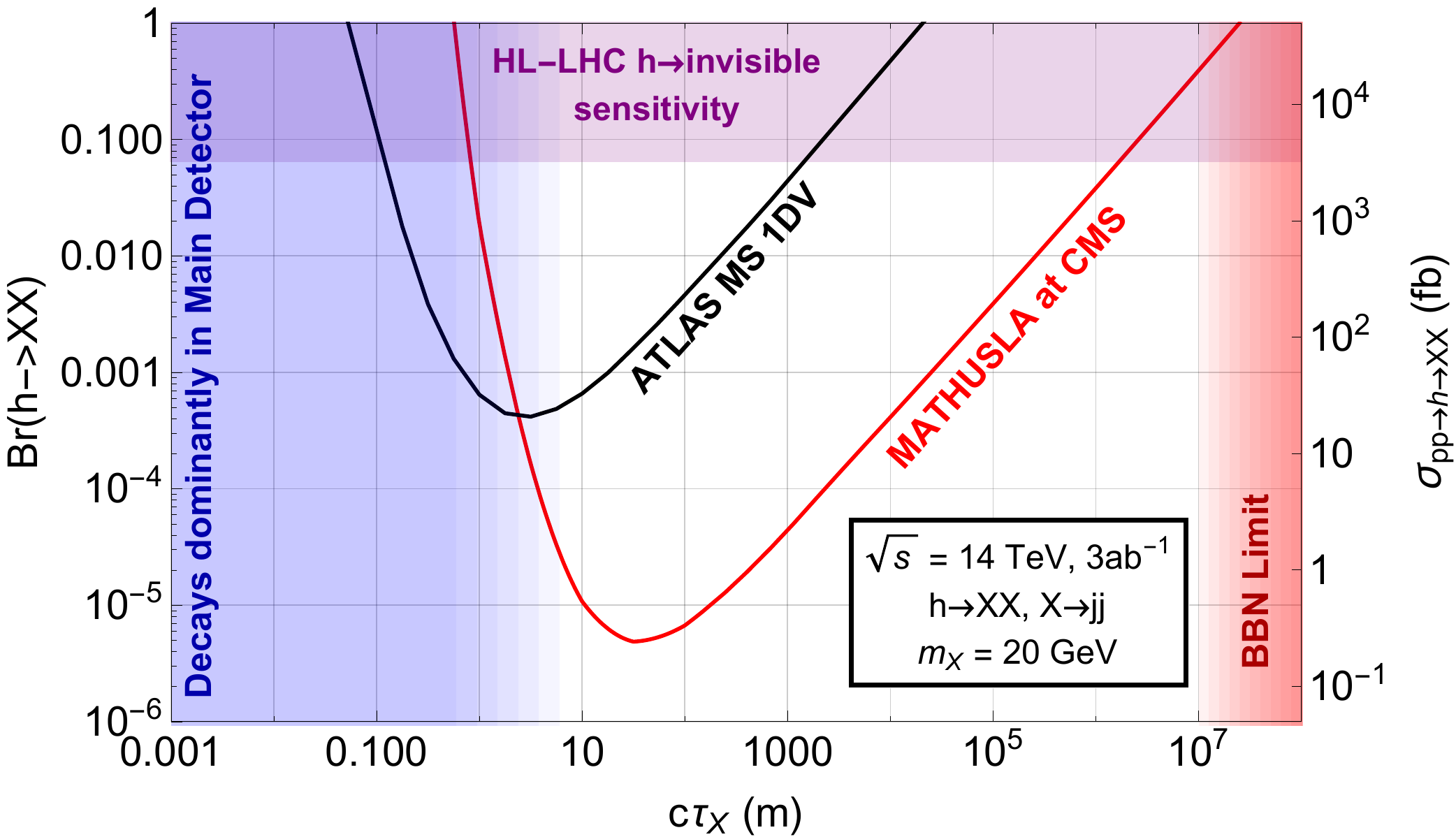}
    \caption{
    Red curve: MATHUSLA@CMS sensitivity (4 observed events) for LLPs of mass $m_X = 20 \gev$ produced in exotic Higgs decays. Black curve: reach of ATLAS search for a single hadronic LLP decay in the Muon System at the HL-LHC~\cite{Coccaro:2016lnz}.
    }
    \label{fig:sensitivity_higgs}
\end{figure}

\begin{figure}
    \centering
    \includegraphics[width=0.9\textwidth]{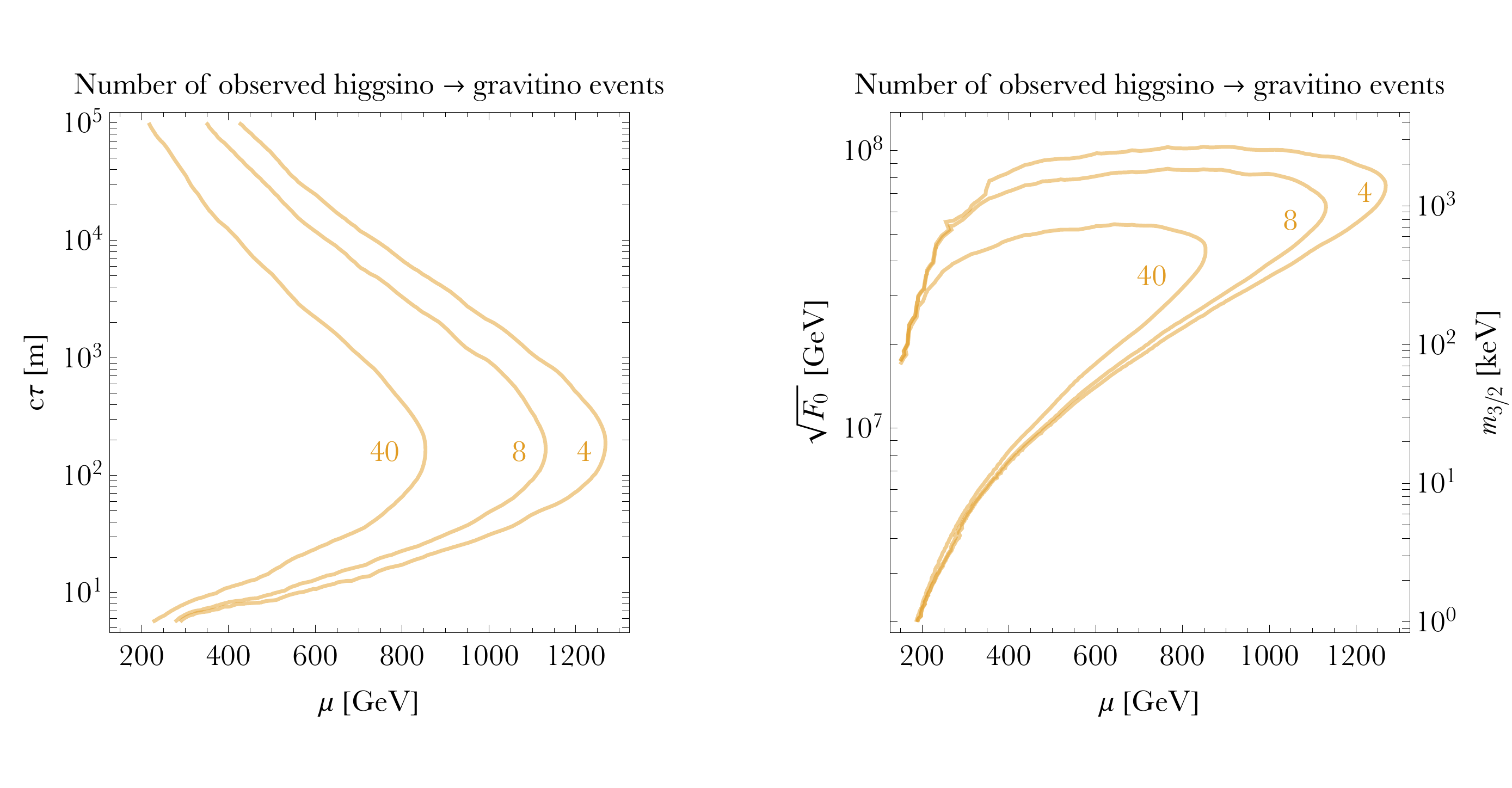}
    \vspace*{-5mm}
    \caption{
    Number of $\tilde H \to \tilde G + (Z,h)$ events that MATHUSLA@CMS could observe from electroweak production of higgsinos at the HL-LHC with an integrated luminosity of 3 ab$^{-1}$. \emph{Left:} higgsino mass $\mu$ versus lifetime $c \tau$ in meters. \emph{Right:} higgsino mass $\mu$ versus the SUSY breaking scale as parametrized by $\sqrt{F}$ in GeV (label on left axis) or gravitino mass $m_{3/2}$ in keV (label on right axis). In a wide swath of parameter space with higgsino lifetimes ranging from smaller than $10$~m to larger than $10^5$~m, MATHUSLA could provide a discovery of new physics with electroweak cross-sections for which the HL-LHC would fail to discover new physics. 
    }
    \label{fig:sensitivity_higgsinos}
\end{figure}

The physics case for LLP searches in general and at MATHUSLA in particular was discussed in great detail in the physics case white paper~\cite{Curtin:2018mvb} and subsequent studies~\cite{Bauer:2018uxu,
Ibarra:2018xdl,
Nelson:2018iuc,
Curtin:2018ees,
Berlin:2018jbm,
Dercks:2018eua,
Ariga:2018uku,
Demidov:2018odn,
Beacham:2019nyx,
Wang:2019orr,
deNiverville:2019xsx,
Serra:2019omd,
Das:2019fee,
Boiarska:2019vid,
Chun:2019nwi,
No:2019gvl,
Krovi:2019hdl,
Bauer:2019vqk,
Wang:2019xvx,
Jana:2019tdm,
Hirsch:2020klk,
Banerjee:2020kww,
Kling:2020mch,
Barron:2020kfo,
Dreiner:2020qbi}.
All the sensitivity estimates in~\cite{Curtin:2018mvb} and~\cite{Bauer:2018uxu,
Ibarra:2018xdl,
Nelson:2018iuc,
Curtin:2018ees,
Berlin:2018jbm,
Dercks:2018eua,
Ariga:2018uku,
Demidov:2018odn,
Beacham:2019nyx,
Wang:2019orr,
deNiverville:2019xsx,
Serra:2019omd,
Das:2019fee,
Boiarska:2019vid,
Chun:2019nwi,
No:2019gvl,
Krovi:2019hdl,
Bauer:2019vqk,
Wang:2019xvx,
Jana:2019tdm,
Hirsch:2020klk,
Banerjee:2020kww,
Kling:2020mch,
Barron:2020kfo,
Dreiner:2020qbi}
assumed the MATHUSLA200 benchmark geometry from the original Letter of Intent~\cite{Alpigiani:2018fgd}, which is identical to the original proposal~\cite{Chou:2016lxi}: a 200~m $\times$ 200~m $\times$ 20~m decay volume, with 100~m displacement both horizontally and vertically from the LHC interaction point. 
The much more realistic MATHUSLA@CMS benchmark geometry introduced in this note is significantly smaller, with a partially excavated decay volume of 
 a 100~m $\times$ 100~m $\times$ 25~m, displaced 70~m horizontally and 60~m vertically from the CMS interaction point.
We now show updated sensitivity curves of MATHUSLA@CMS for several LLP benchmark models at the LHC: exotic Higgs decays, long-lived Higgsinos, as well SM+S (light scalar mixing with the Higgs) and RHN (Right-Handed Neutrinos). This demonstrates that the sensitivity is essentially identical to the old MATHUSLA200 benchmark, since the reduced distance from the IP makes up for the smaller decay volume. Therefore, all of the MATHUSLA200 sensitivity estimates apply for the updated MATHULSA@CMS benchmark almost verbatim.\footnote{The only significant difference is that MATHUSLA@CMS has an improved sensitivity for shorter LLP lifetimes compared to MATHUSLA200, again due to the smaller distance to the IP.}

\begin{figure}
    \centering
        \hspace*{-23mm}
    \begin{tabular}{c}
     \includegraphics[width=0.6\textwidth]{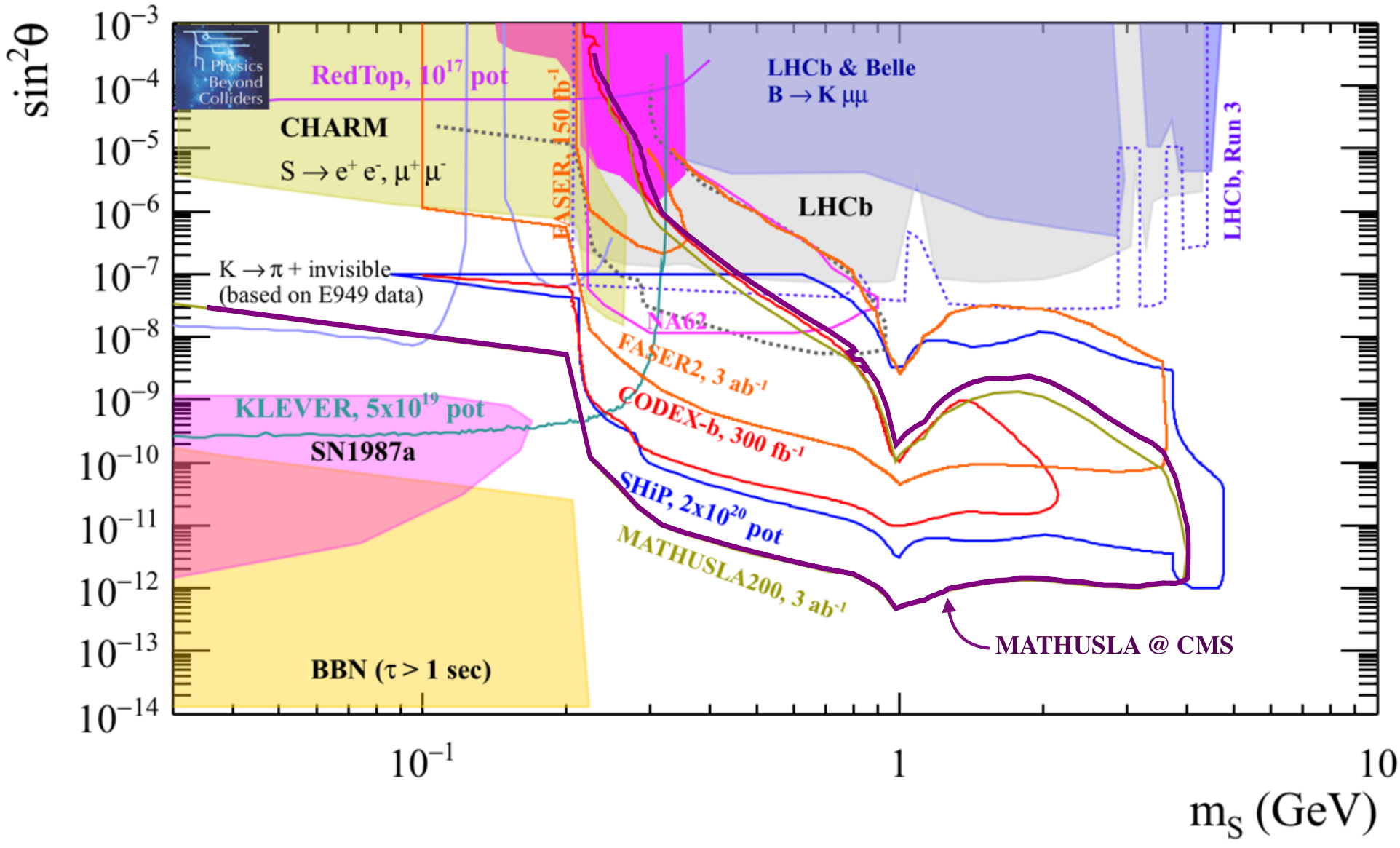}
     \vspace*{-5mm}
    \\
    (a)
    \vspace*{2mm}
    \\
    \begin{tabular}{cc}
      \includegraphics[width=0.6\textwidth]{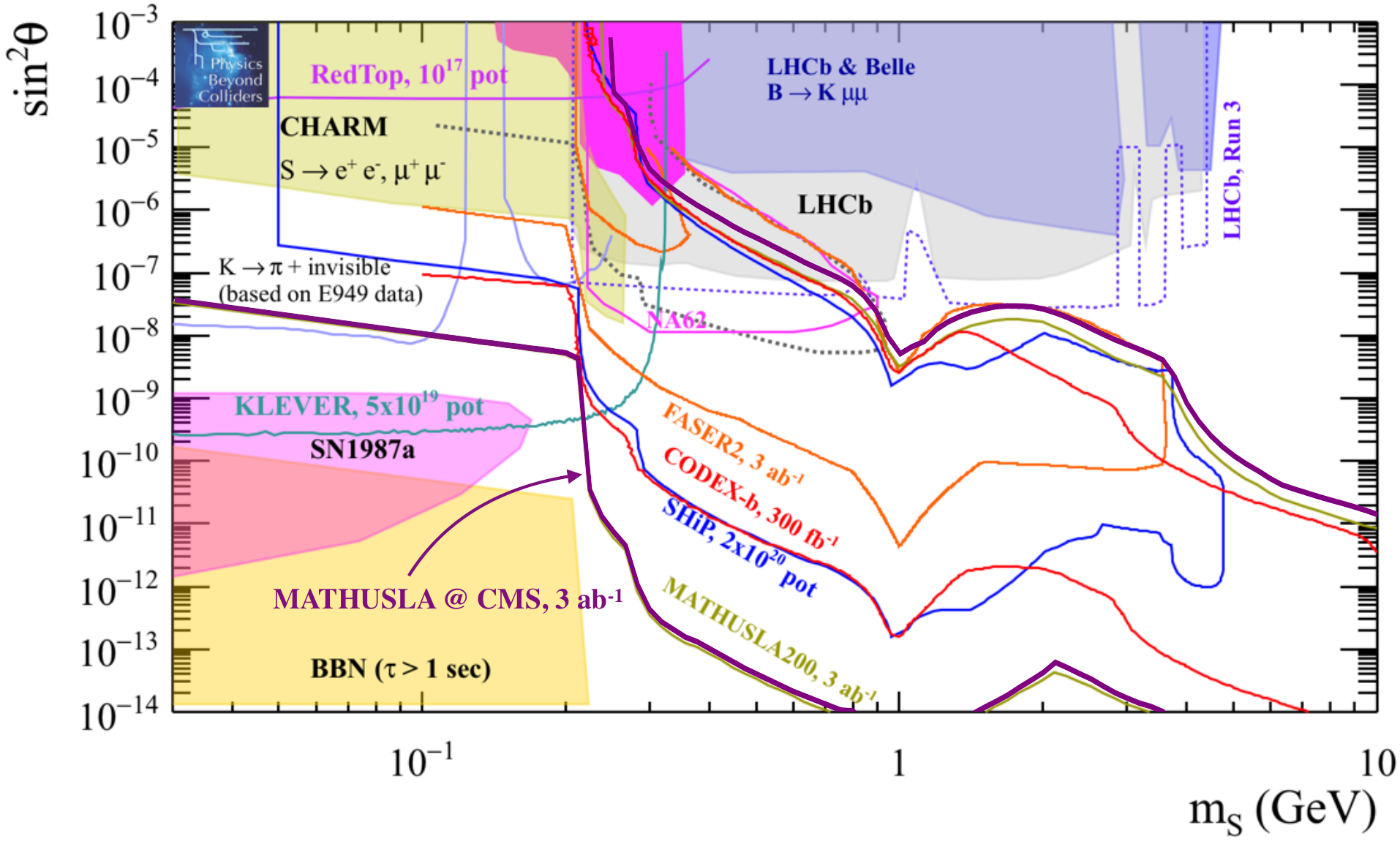}  
      &
        \includegraphics[width=0.6\textwidth]{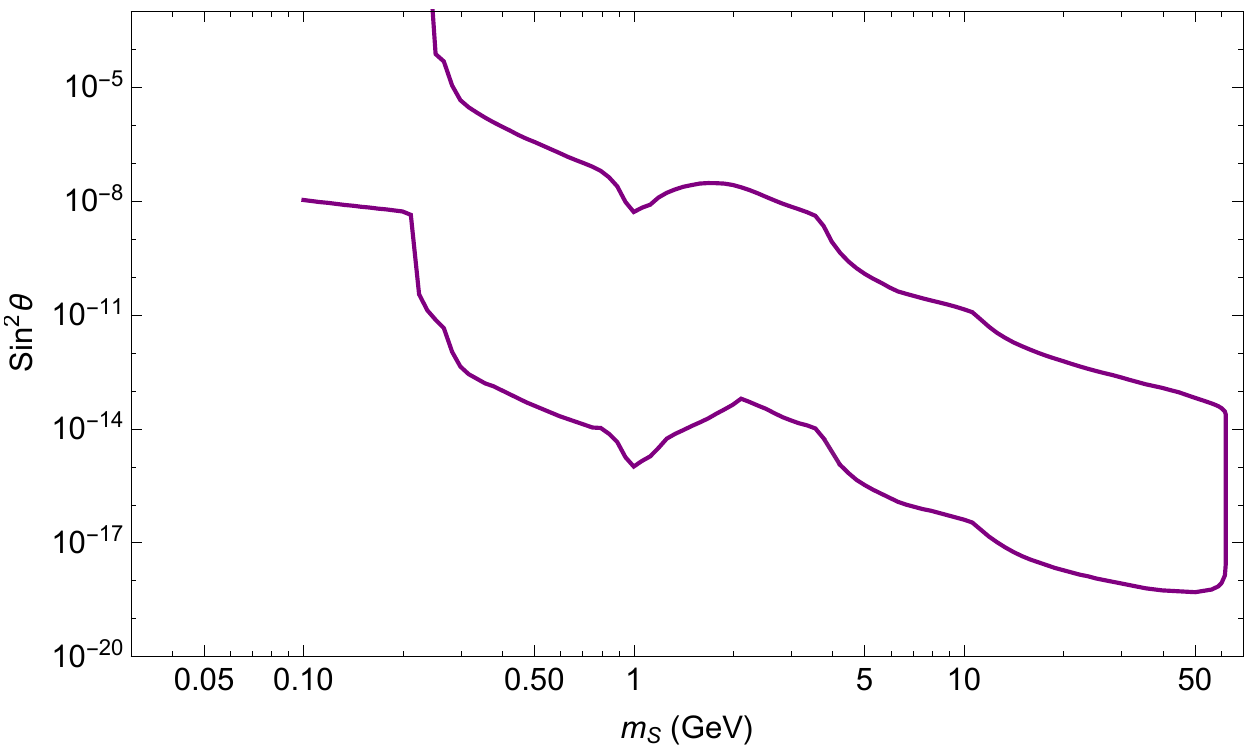}    \vspace*{-0mm}
    \\
    (b) & (c)
    \end{tabular}
    \end{tabular}
    \caption{
    Purple curves: sensitivity of MATHUSLA@CMS for a singlet scalar LLP $s$ mixing with Higgs mixing angle $\theta$. (a) Assuming production in exotic $B$, $D$, $K$ meson decays only. (b) Assuming additional production in exotic Higgs decays with $\mathrm{Br}(h\to ss) = 0.01$. Figures (a) and (b) are reproduced from the PBC BSM Working Group report~\cite{Beacham:2019nyx} with the purple MATHUSLA@CMS curves added. This  shows sensitivity of various other existing and proposed experiments, as well as the old MATHUSLA200 benchmark estimates (yellow curves). (c) Same scenario as (b) but showing the entire MATHUSLA sensitivity due to $h\to ss$ decays.}
    \label{fig:sensitivity_SMS}
\end{figure}

\begin{figure}
    \centering
    \hspace*{-23mm}
     \begin{tabular}{cc}
     \includegraphics[width=0.6\textwidth]{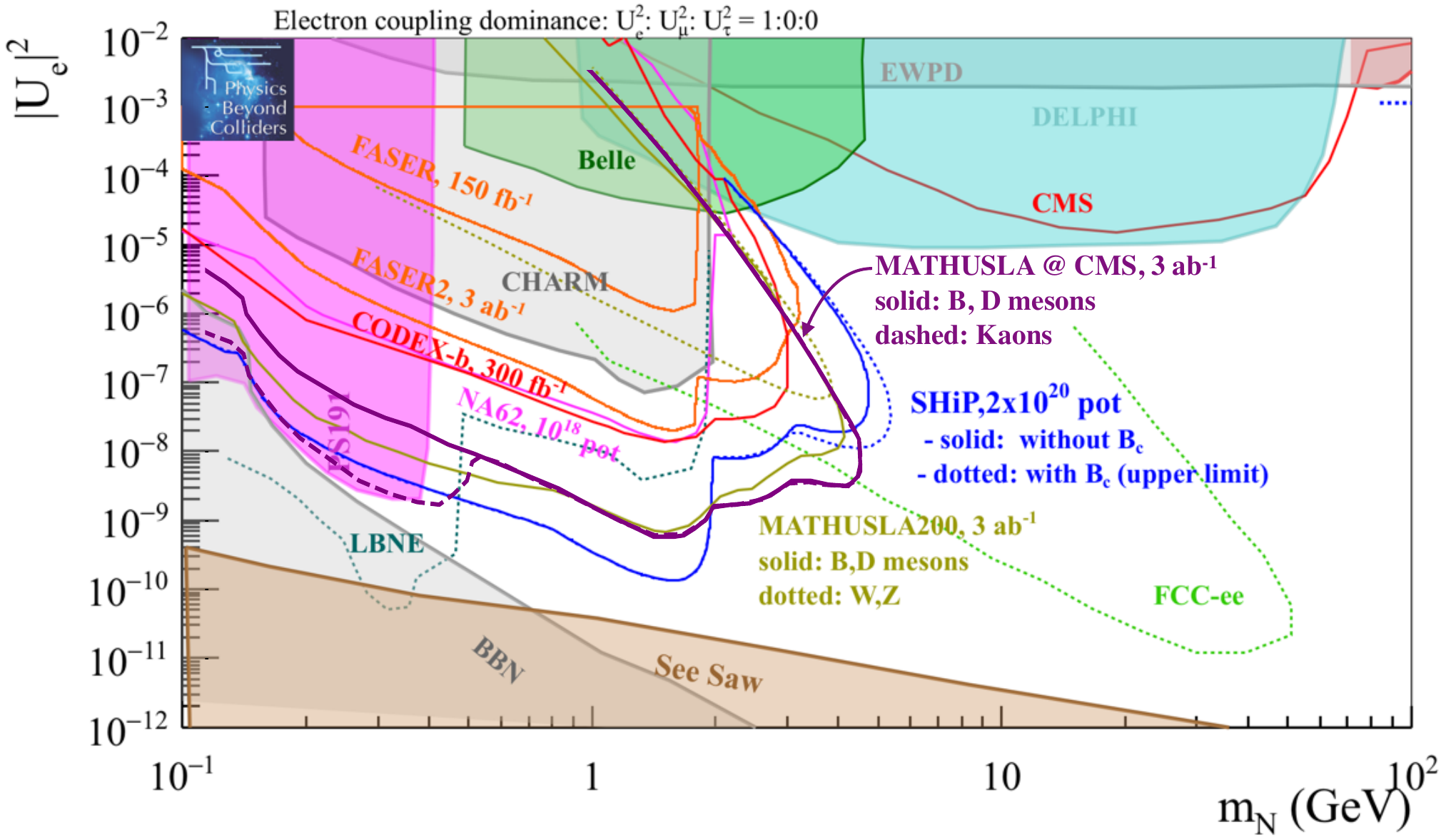}
     &
      \includegraphics[width=0.6\textwidth]{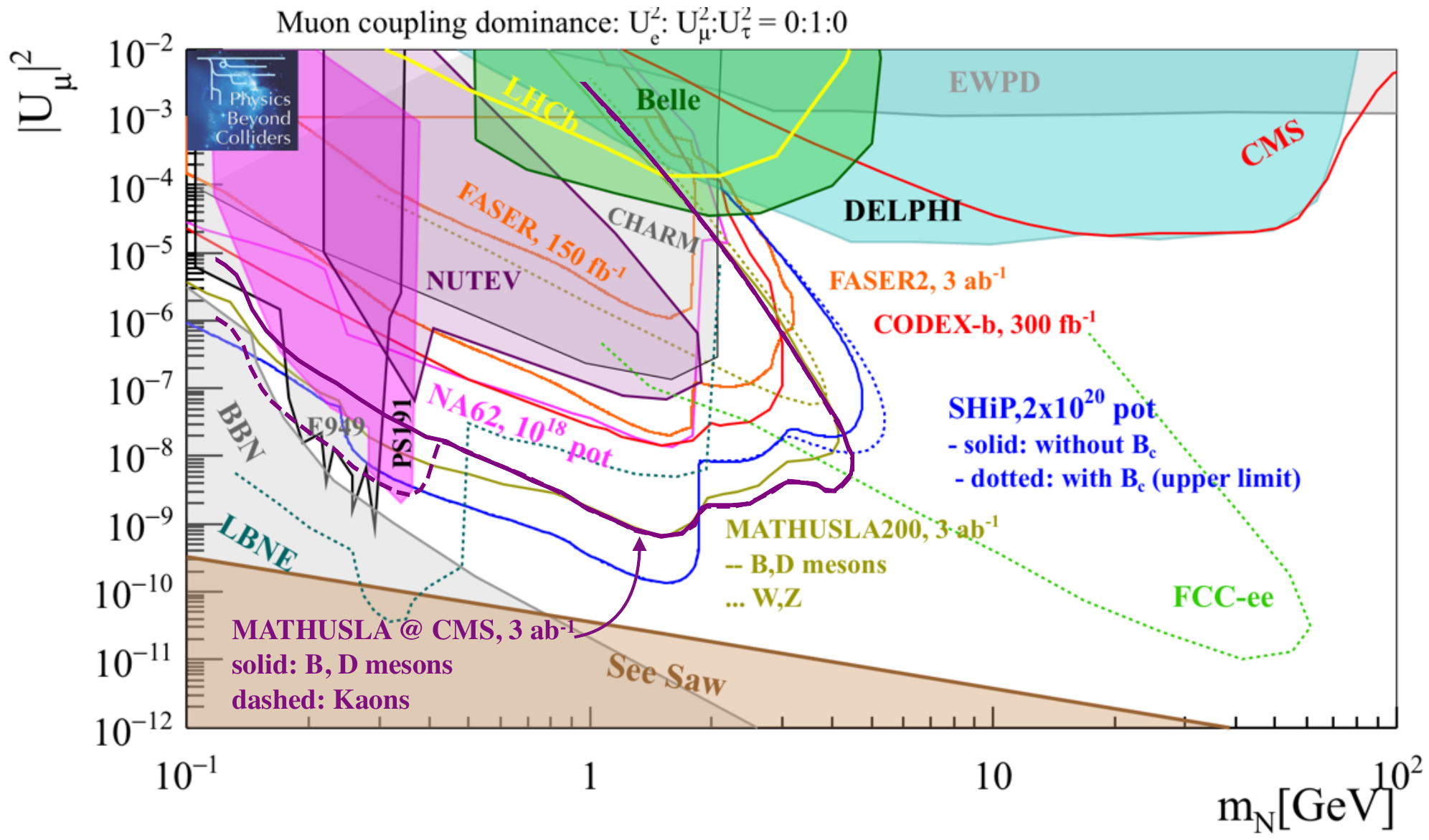}  
      \\
      (a) & (b)
      \\
        \includegraphics[width=0.6\textwidth]{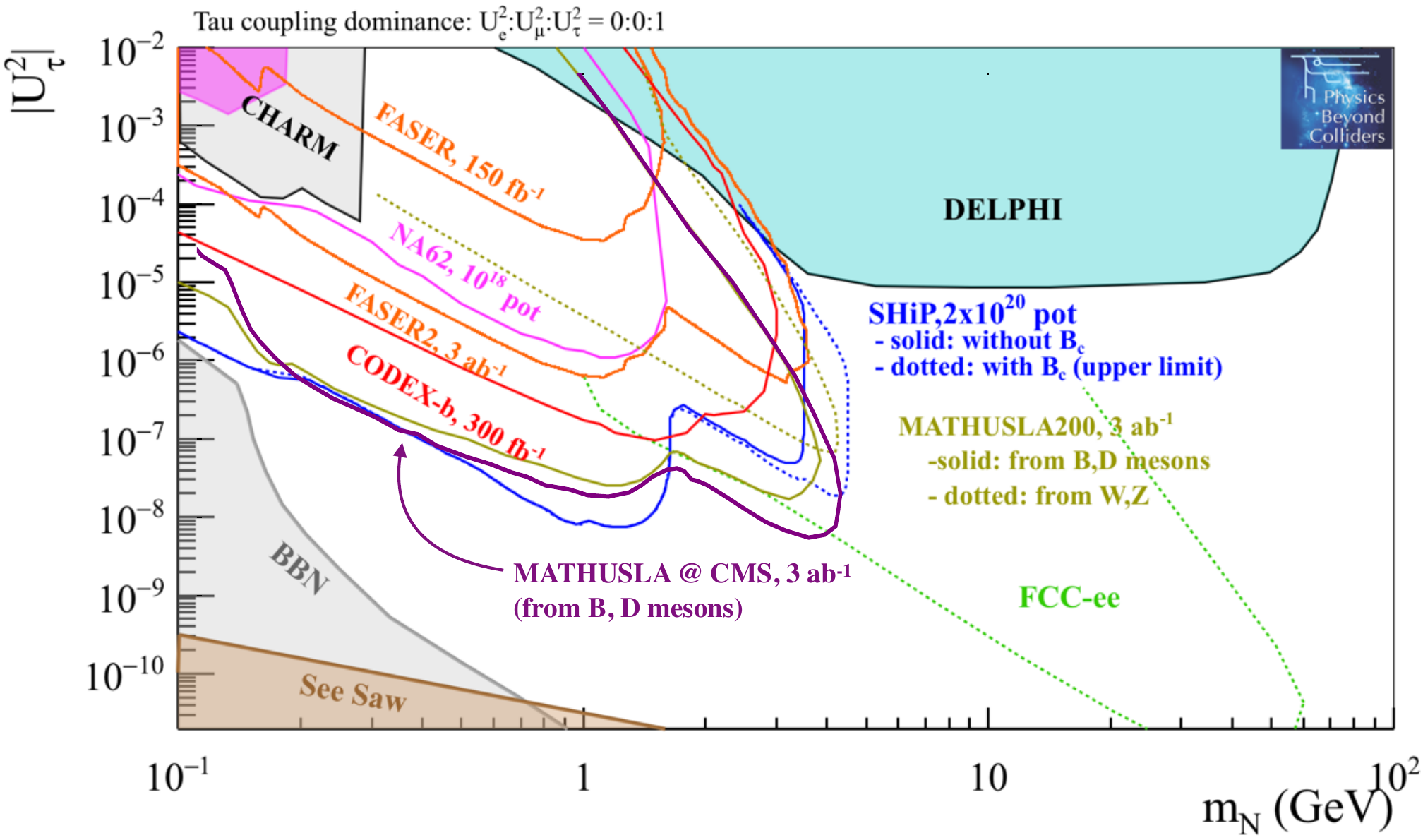}    \vspace*{-0mm}
    &
      \includegraphics[width=0.6\textwidth]{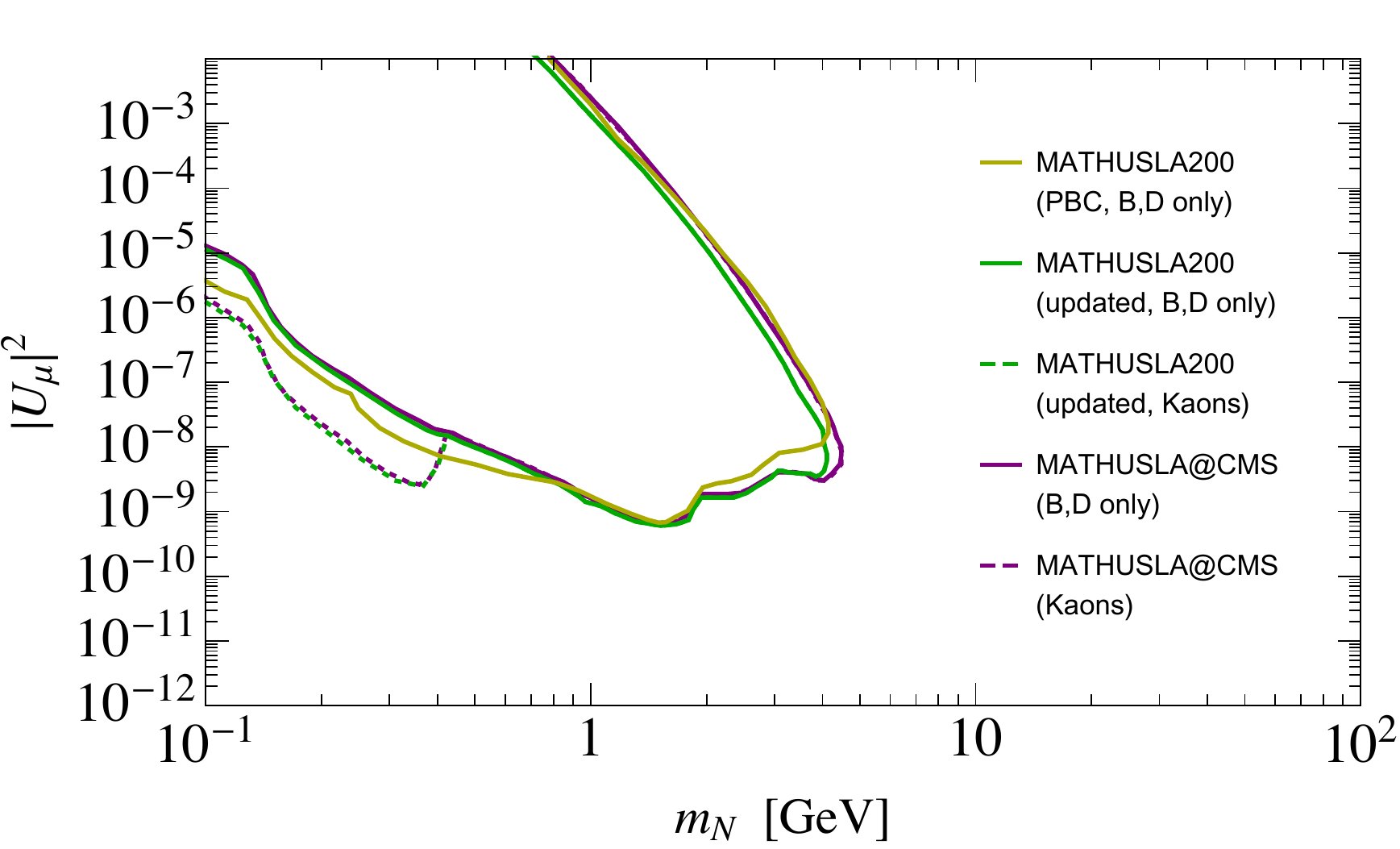}
    \\
    (c) & (d)
    \end{tabular}
    \caption{
    Purple curves: sensitivity of MATHUSLA@CMS for a Heavy Neutral Lepton LLPs produced in $B, D$ (solid) and $K$ decays (dashed) for HNLs that mix predominantly with electron (a), muon (b) or tau (c) active neutrinos. These figures are reproduced from the PBC BSM Working Group report~\cite{Beacham:2019nyx} with the purple MATHUSLA@CMS curves added. This  shows sensitivity of various other existing and proposed experiments, as well as the old MATHUSLA200 benchmark estimates (yellow curves). (d) For the muon-dominated mixing scenario, comparison of the old MATHUSLA200 sensitivity estimate in the PBC and MATHUSLA physics case white papers~\cite{Curtin:2018mvb,Bondarenko:2018ptm} (dark yellow) to the MATHUSLA200 (green) and MATHUSLA@CMS (purple) sensitivity computed using the latest results in~\cite{Bondarenko:2018ptm} (blue). 
    The dark yellow and purple curves are the same as in plot (b). 
    The small difference between the green and purple curves demonstrates that the new MATHUSLA geometry has very little effect on sensitivity, and the difference between the yellow and purple curves in (a)-(c) is mainly due to the improved HNL production and decay calculation.
    }
    \label{fig:sensitivity_HNL}
\end{figure}

Note that the sensitivity estimates we present here, like those of~\cite{Curtin:2018mvb}, assume perfect  detection efficiency as long as the LLP decays in the decay volume, and assume zero backgrounds after the rigorous geometric DV reconstruction cuts have been applied. 
Both the zero-background assumption (see new studies in Section~\ref{s.backgrounds}) and perfect reconstruction assumption (see preliminary reconstruction studies in LOI~\cite{Alpigiani:2018fgd}) are good approximations for LLPs decaying into a high multiplicity of final states, which is the case for MATHUSLA's most important physics target, hadronically decaying LLPs with masses in the $\mathcal{O}(10 \gev) - \mathcal{O}(100 \gev)$ range. 
The extent to which these ideal assumptions hold for leptonically decaying or very light LLPs with masses $\lsim$ GeV is the subject of ongoing study by the MATHUSLA collaboration, and also depends on details of the final detector design.

We first consider examples of weak- or TeV-scale LLPs produced at the LHC.
Figure~\ref{fig:sensitivity_higgs} shows the sensitivity to hadronically decaying LLPs produced in exotic Higgs decays, which arises in a large variety of new physics scenarios~\cite{Curtin:2013fra}, including solutions to the Hierarchy Problem like  Neutral Naturalness~\cite{Chacko:2005pe,Burdman:2006tz,Craig:2015pha,Curtin:2015fna}.
The LLP cross section sensitivity on the right axis approximately applies to most weak-scale LLP production processes~\cite{Curtin:2018mvb}. 
Figure~\ref{fig:sensitivity_higgsinos} shows the reach for meta-stable Higgsinos within Supersymmetry with gauge-mediated SUSY breaking~\cite{Giudice:1998bp,Knapen:2016exe}. 
In both cases, the observable LLP production rate is nearly identical to the sensitivity of the old MATHUSLA200 benchmark~\cite{Chou:2016lxi,Alpigiani:2018fgd}, with slightly improved sensitivity at shorter lifetimes. 

We now consider two of the benchmark models for low-mass LLP that were also studied by the Physics Beyond Colliders (PBC) working group~\cite{Beacham:2019nyx}. 
Figure~\ref{fig:sensitivity_SMS} shows the sensitivity for a singlet scalar LLP that has a tiny mixing angle $\theta$ with the Higgs boson (for details see~\cite{Evans:2017lvd,Beacham:2019nyx}). The sensitivity is again nearly identical to MATHUSLA200. Part (b), where an exotic Higgs decay branching fraction of $\mathrm{Br}(h \to ss) = 0.01$ is assumed as an additional LLP production process, demonstrates the advantage gained by the high LHC energy even when searching for very light LLPs, since they can be produced in high-scale processes that are kinematically suppressed at intensity frontier experiments. 
Also shown are the expected sensitivities of other proposed experiments like FASER~\cite{Ariga:2018zuc}, CODEX-b~\cite{Aielli:2019ivi}, and SHiP~\cite{Alekhin:2015byh}. 

Figure~\ref{fig:sensitivity_HNL} (a) - (c) shows MATHUSLA's reach for Heavy Neutral Leptons (HNL) that dominantly mix with only electron, muon or tau active neutrinos.  
These estimates are slightly improved compared to the previous results in~\cite{Curtin:2018mvb,Beacham:2019nyx}, since the production and decay rate calculations have been updated to agree with the latest results in~\cite{Bondarenko:2018ptm}, also adopted by e.g.~\cite{Coloma:2020lgy}. 
The effect of this improved calculation for the old MATHUSLA200 benchmark geometry can be seen in Figure~\ref{fig:sensitivity_HNL} (d) by noting the small differences between the dark yellow and green curves. 
The effect of the new MATHUSLA@CMS geometry is then extremely minor, same as for the other LLP production modes, as can be seen by comparing the green to the purple curve. 
Additionally, the production of HNLs in Kaon decays is now taken into account (dashed curves). This is denoted separately, since the resulting HNLs are relatively soft and MATHUSLA's ability to reconstruct such light, soft LLPs depends on the final detector design.

%%%%%%%%%%%%%%%%%%%%%%%%%%%%%%%%%%
%%%%%%%%%%%%%%%%%%%%%%%%%%%%%%%%%%
%%%%%%%%%%%%%%%%%%%%%%%%%%%%%%%%%%
\subsection{Dark Matter}
\label{s.DM}

It is perhaps not commonly appreciated that LLP searches are crucial to complete the search program for DM.
In models like Freeze-In DM (FIDM)~\cite{Hall:2009bx, No:2019gvl}, inelastic DM (iDM)~\cite{TuckerSmith:2001hy, Izaguirre:2015zva}, co-annihilating DM~\cite{DAgnolo:2018wcn} or co-scattering DM~\cite{Aielli:2019ivi}, the relic abundance of the stable DM candidate is determined by the properties of an LLP in the thermal plasma of the Big Bang.
This LLP carries the same quantum number which stabilizes DM, and decays into DM + SM final states.
The DM particle itself could be almost completely sterile, precluding a direct detection signal, and production of the parent LLP at colliders could then be the only way to produce and observe DM. 

The mass of the LLP can easily exceed the GeV scale,  meaning they can only be produced in the high-energy collisions of the LHC,
while  the connection between the LLP properties and the DM relic abundance often pushes the necessary LLP searches into the long-lifetime regime. 
All of this means that MATHUSLA  has a unique and invaluable role to play in the search for Dark Matter, which we demonstrate in this section by briefly reviewing several recent studies from the literature.\footnote{While we focus on the DM reach from LLP searches, there are also much more exotic possibilities. For example, MATHUSLA could at as a \emph{direct detection experiment} for strongly-interacting very heavy DM, showing up as a very slow downward track with multiple hits in each tracker element~\cite{Bramante:2018qbc}.} (All of these studies used the original MATHUSLA200 benchmark, but as we demonstrated in the last section, those results also apply to the updated MATHUSLA@CMS geometry.)

We first consider Freeze-In DM~\cite{Hall:2009bx, No:2019gvl}.
In FIDM, the DM candidate $X$ is essentially sterile, with tiny or non-existent direct coupling to SM particles. 
$X$ therefore never reaches equilibrium with the SM bath during the Big Bang, and its abundance, assumed to be negligibly small after reheating at the end of inflation, cannot be set by thermal freeze-out.
However, if there is a parent particle $A$ in thermal equilibrium with the plasma that has a long-lived decay
\begin{equation}
A \to B_{SM} X \ ,
\end{equation}
where $B_{SM}$ is some SM state, then the $X$ abundance can be gradually populated while $A$ is relativistic during the radiation-dominated era,\footnote{Assuming standard cosmology, see~\cite{Co:2015pka,DEramo:2017ecx} for the impact of modified thermal histories}. Most of the $X$ particles are produced around $T \sim m_A/3$, just before $A$ disappears from the plasma due to its efficient annihilation into SM particles. This is referred to as DM ``freeze-in''.\footnote{Freeze-in scenarios can also utilize suppressed annihilations of particles in the SM bath to produce $X$, for a recent review see e.g.~\cite{Alexander:2016aln}.}

\begin{figure}
    \centering
    \hspace*{-15mm}
    \begin{tabular}{m{0.4\textwidth}m{0.42\textwidth}
    m{0.3\textwidth}}
    \includegraphics[width=0.4 \textwidth]{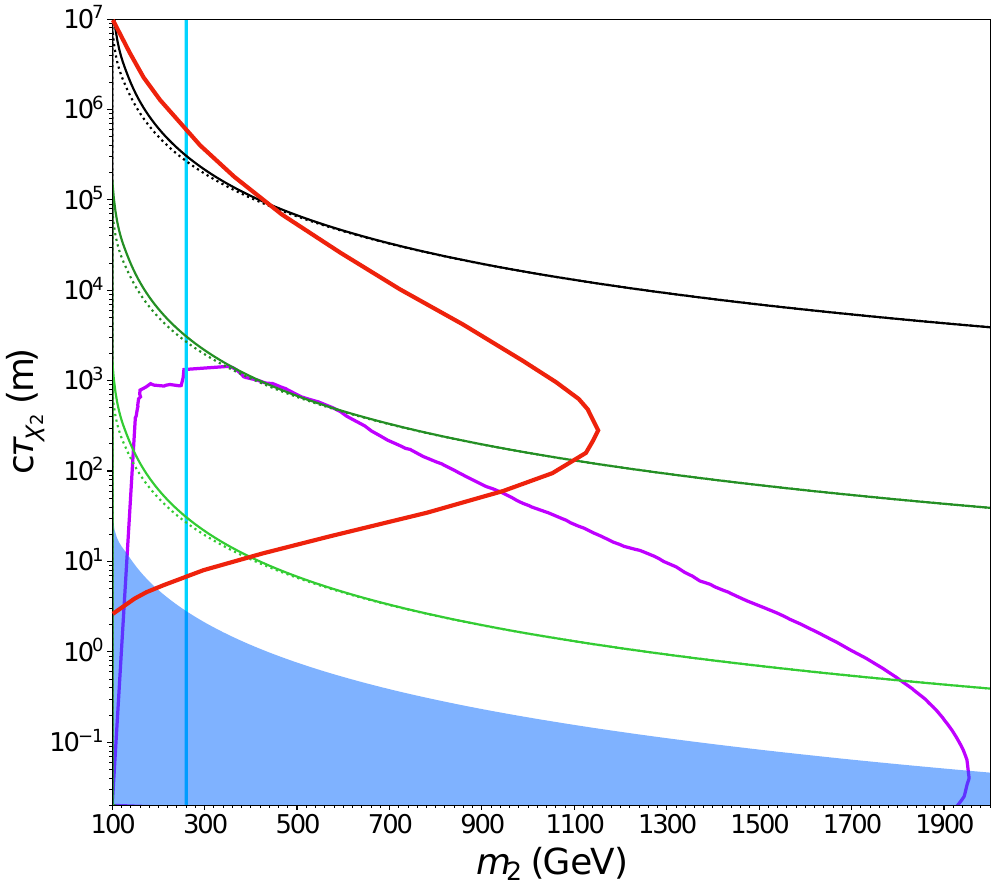}
     &
    \includegraphics[width=0.44 \textwidth]{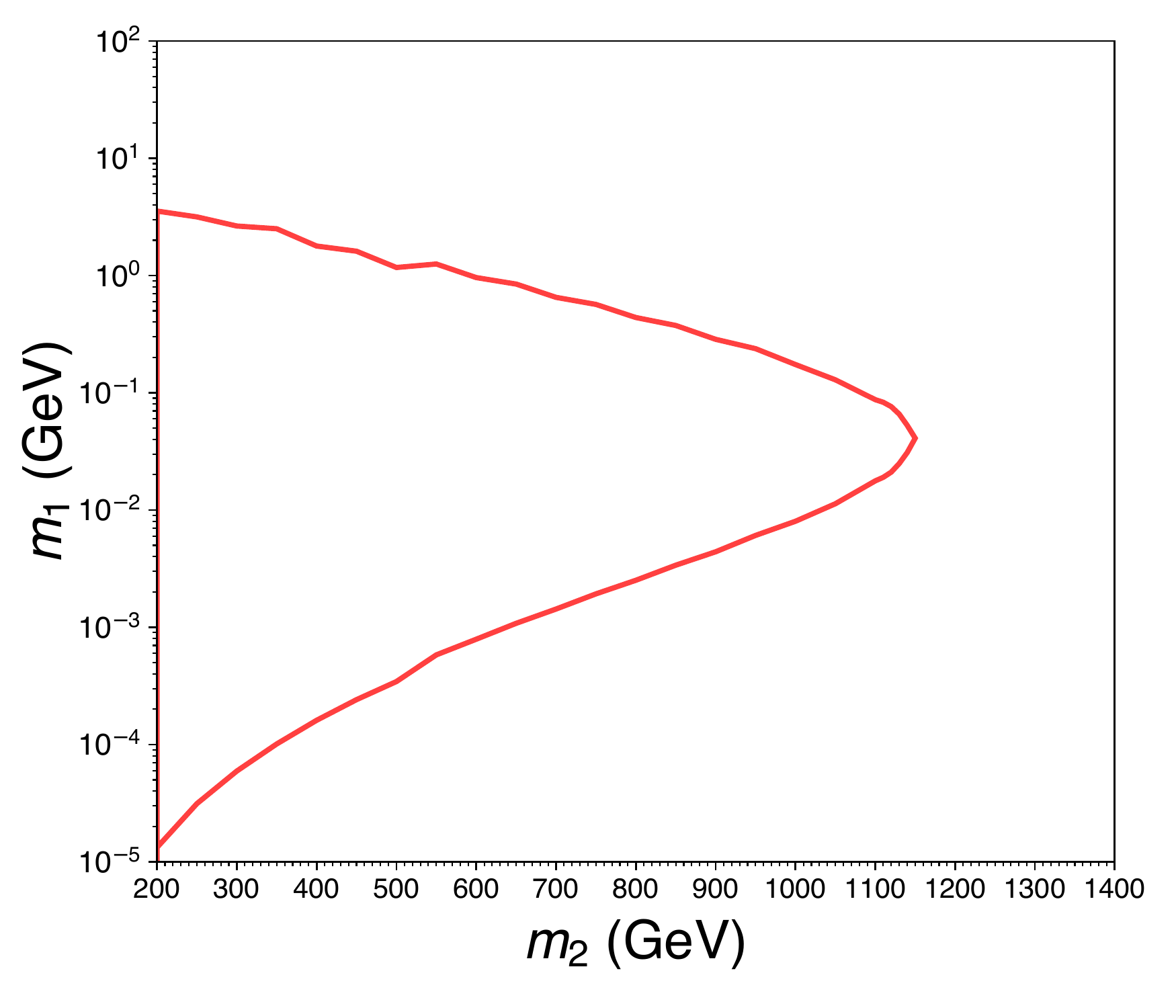}
    &
     \includegraphics[width=0.3 \textwidth]{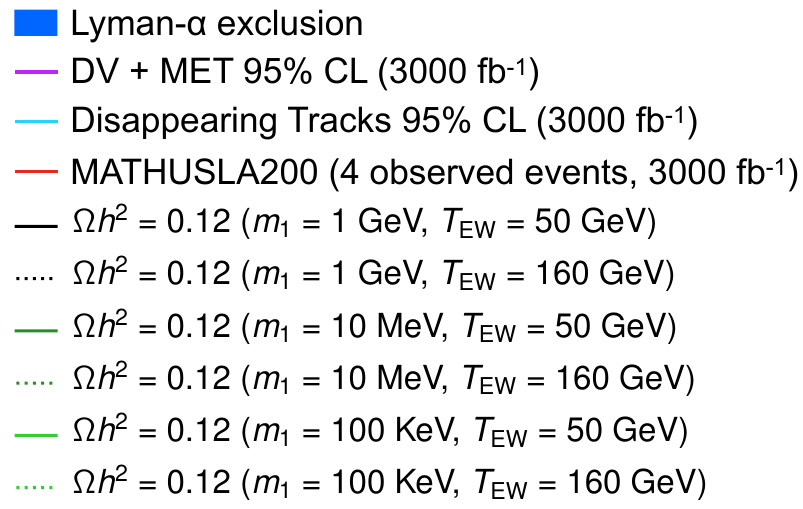}
    \end{tabular}
    \caption{
    \emph{Left:} HL-LHC reach of MATHUSLA (red, 4 observed events) and ATLAS/CMS searches for DV + MET (purple) and disappearing tracks (cyan)
    in the mass-lifetime plane of the parent LLP $\chi_2$ in our simplified FIDM model. The blue region excluded from Large-Scale Structure constraints derived from Lyman-$\alpha$ forest observations. 
    The solid and dotted near-horizontal lines denote mass-lifetime relationship required for a 1 GeV, 10 MeV or 100 KeV DM particle $\chi_1$ to have the observed relic abundance, for different temperatures of the electroweak phase transition.
   Figure adapted from~\cite{No:2019gvl} and~\cite{Curtin:2018mvb}.
   \emph{Right:} reach of MATHUSLA (4 observed events) in the LLP-DM mass plane. This figure is adapted from~\cite{Curtin:2018mvb} and assumes the mass-lifetime relationship of the earlier analysis which differs slightly from the newer analysis~\cite{No:2019gvl}, but still illustrates the reach of MATHUSLA in the DM mass plane.
    }
    \label{fig:FIDM}
\end{figure}

The lifetime and mass of the parent particle $A$ directly set the DM abundance, and since it has sizable coupling to the SM, $A$ is an LLP that can be produced at colliders. 
A simplified model of FIDM, considered in~\cite{Curtin:2018mvb} and further analyzed in~\cite{No:2019gvl}, serves to illustrate the DM reach of LLP searches. 
It consists of one electroweak doublet Dirac fermion $\psi$ and one singlet Dirac fermion $\chi$ with a higgs couping $y_\chi \bar \psi H \chi$. In the regime $m_\psi > m_\chi$ and $y_\chi \ll 1$, the two fermions acquire a small mixing after electroweak symmetry breaking. The $\psi$-dominated mass eigenstate $\chi_2$ with mass $m_2$ then acts as the parent LLP with sizable SM couplings, decaying to the $\chi$-dominated mass eigenstate $\chi_1$ with mass $m_1$, which is the FIDM candidate. This model is similar to the singlet-doublet DM model considered in \cite{Calibbi:2018fqf,Calibbi:2015nha}, and serves as a close analogue of the Higgsino-Axino system~\cite{Covi:1999ty, Co:2015pka, Covi:2001nw, Co:2016fln} or a feebly interacting Higgsino-Bino system. Figure~\ref{fig:FIDM} shows the reach of MATHUSLA in either the mass-lifetime plane of the LLP (with near-horizontal curves showing where DM of varying mass has the correct relic abundance), or the LLP-DM mass plane. Also shown is the projected reach of disappearing track and DV+MET searches at ATLAS/CMS for the HL-LHC. 
DM masses above the MeV scale imply the long-lifetime regime for the parent LLP, making MATHUSLA the most sensitive experiment to probe large regions of FIDM parameter space

Another example of this very direct LLP-DM connection is inelastic DM~\cite{TuckerSmith:2001hy, Izaguirre:2015zva}, where the stable DM state $\chi_1$ can only interact with the SM by turning into a slightly heavier and meta-stable dark state $\chi_2$. 
This means direct detection experiments can only detect iDM with a mass splitting smaller than the nuclear recoil energy $\lsim 100 \kev$, and larger mass splittings have to be probed at colliders.

\begin{figure}
    \centering
    \begin{tabular}{cc}
    \includegraphics[width=0.5\textwidth]{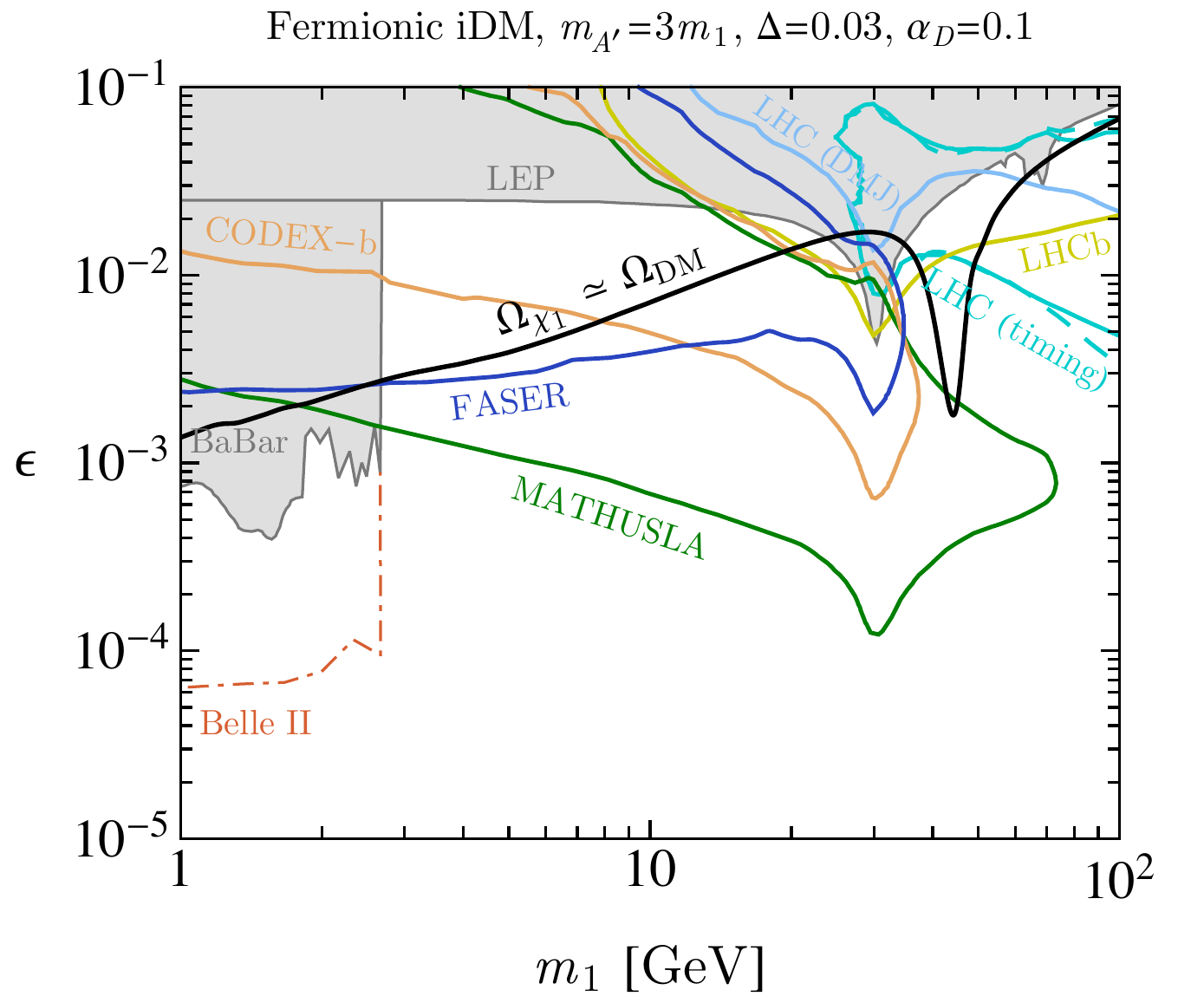}
    &
     \includegraphics[width=0.5\textwidth]{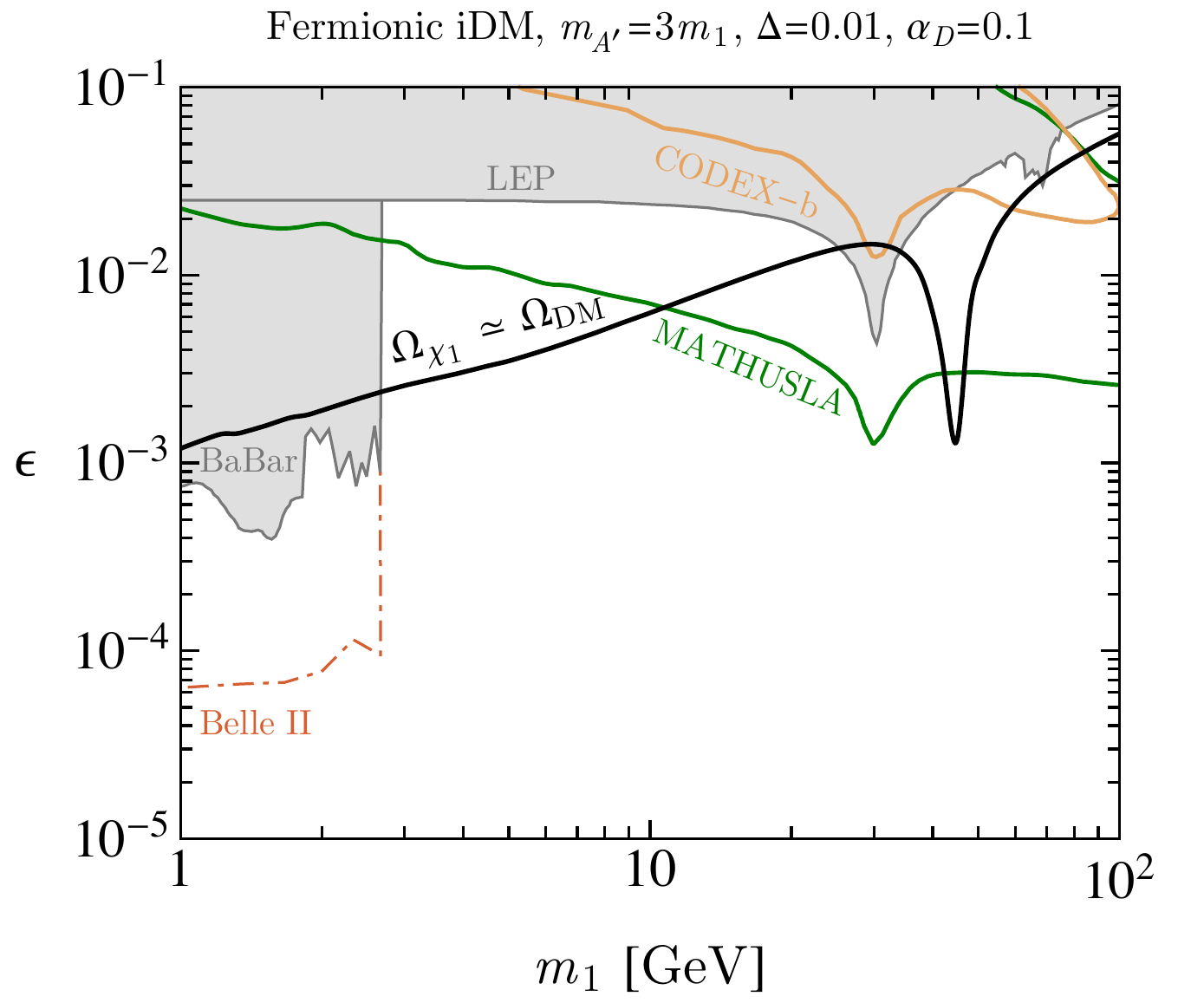}
     \end{tabular}
    \caption{
    Reach of MATHUSLA and other LHC experiments and searches for iDM with a dark photon  of mass $m_{A'}$ that has kinetic mixing $\epsilon$ with the SM photon, and mass splittings $\Delta$ in the percent range. The black curve indicates where thermal o-annihilations $\chi_2 \chi_1 \to A' \to f \bar f$ to SM fermions give the observed DM relic density. Figure taken from~\cite{Berlin:2018jbm}.
    }
    \label{fig:iDM}
\end{figure}

The recent study Ref.~\cite{Berlin:2018jbm}
examined a simple iDM model with a Dirac pair of two dark Weyl fermions oppositely charged under a dark broken $U(1)_D$ that has mass $m_{A'}$ and kinematic mixing $\epsilon$ with the SM photon. In addition to the Dirac mass between the two fermions, symmetry breaking generates small Majorana mass terms. This results in two mass eigenstates $\chi_{1,2}$ with small mass splitting $\Delta = (m_2 - m_1)/m_1 \lsim 0.1$ and an off-diagonal coupling $i e_D A'_\mu \bar \chi_1 \gamma^\mu \chi^2$ to the dark photon. 
For $m_{A'} > \Delta \cdot m_1$, $\chi_2$ is an LLP that decays to a DM particle $\chi_1$ and two SM fermions via an off-shell $A'$ with a lifetime given dominantly by the mass splitting $\Delta$ and the photon portal coupling $\epsilon$. $\chi_1 \chi_2$ pairs can be produced via an $s$-channel dark photon in colliders and fixed target experiments, allowing LLP searches to probe this class of models.
Figure~\ref{fig:iDM} shows the reach of MATHUSLA, FASER, CODEX-b and various LHC searches for dark matter of mass $m_1$ with dark photon kinetic mixing $\epsilon$ for mass splittings in the percent range $\Delta \sim 0.01$.  The black line shows where $\chi_1$ has the right relic density from thermal freeze-out via $\chi_2 \chi_1 \to A' \to \bar f f$ co-annihilations to SM fermions. Note that only the LHC can produce $\chi_1 \chi_2$ pairs at high rate for dark photon masses above $\sim 10 \gev$, demonstrating the importance of the energy frontier for probing even relatively light DM models, and direct detection experiments have no reach due mass splittings around $0.1-1$ GeV. MATHUSLA provides the best reach and is the only way to probe large parts of motivated iDM parameter space where the DM candidate has the observed relic density, in particular for mass splittings below a percent.

iDM and FIDM demonstrate how the properties of an LLP can be directly tied to the observed relic DM abundance. 
It is also possible for LLPs to be motivated or even necessary to realize a given DM mechanism, but without the properties of the LLP directly setting the precise abundance. 
This is the case for Asymmetric DM (ADM)~\cite{Kaplan:2009ag}, where the baryon asymmetry of the universe has a common origin with an asymmetric DM relic abundance. The primordial asymmetry is shared between the visible and dark sectors by higher dimensional operators, which only have to be large enough to establish chemical equilibrium between the dark and visible sectors. 
Realizing this idea within supersymmetric models, for instance, de-stabilizes the Lightest Supersymmetric Particle (LSP) in the visible sector and makes it an LLP with a decay to both visible and DM states. Crucially, since the relic abundance in ADM is not tied to the significant equilibrium interactions of the DM with SM particles, standard probes of DM, such as direct detection, may be insensitive, so that DM production in LLP decays is the only viable discovery channel.
The reach of MATHUSLA for SUSY with unstable LSPs is demonstrated for unstable Higgsinos in Fig.~\ref{fig:sensitivity_higgsinos} and was studied for RPV SUSY in~\cite{Curtin:2018mvb}. TeV-scale LLPs with a wide range of lifetimes can be probed, and depending on the decay channel, the lifetime sensitivity can be several orders of magnitude better than main detector searches.

%%%%%%%%%%%%%%%%%%%%%%%%%%%%%%%%%%%%%%

\begin{figure}
    \centering
    \begin{tabular}{cc}
    \includegraphics[width=0.5\textwidth]{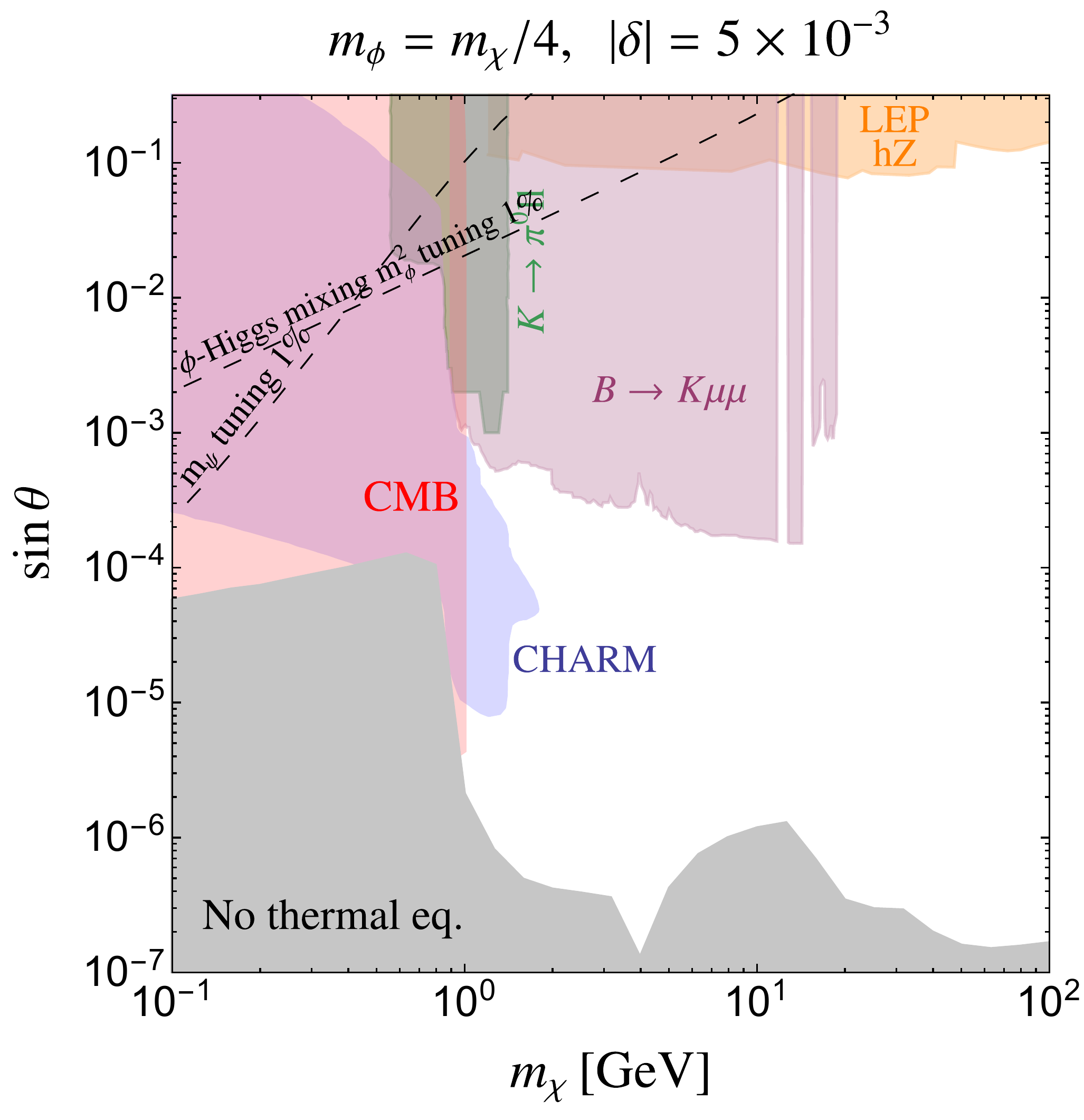}
    &
     \includegraphics[width=0.5\textwidth]{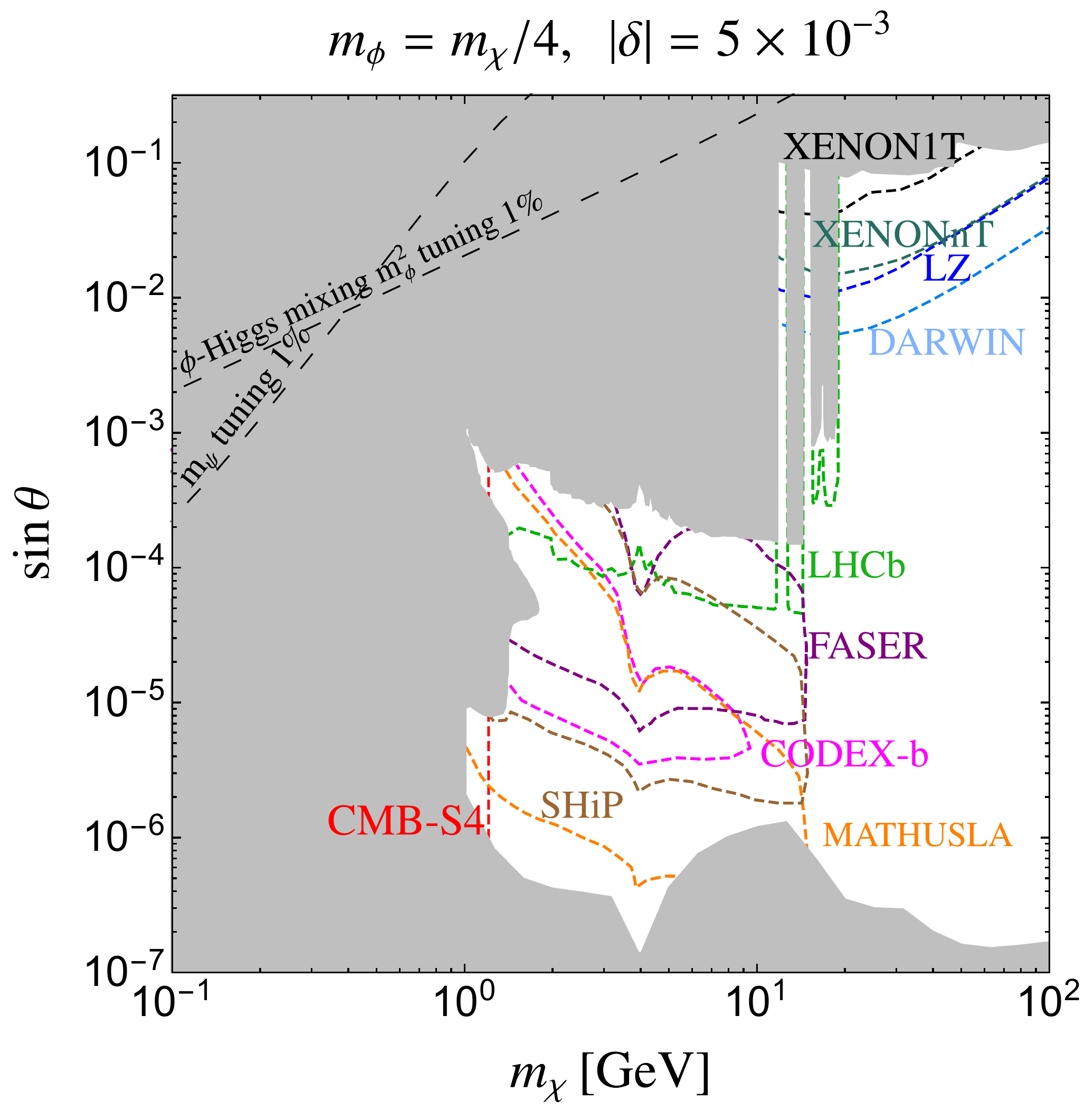}
     \end{tabular}
    \caption{In the co-annihilation scenario studied by~\cite{DAgnolo:2018wcn}, with DM mass $m_\chi$, relative mass splitting to next lightest dark state $\delta$, and light scalar mediator mass $m_\phi$, the left plot shows existing cosmological and collider constraints on the plane of DM mass and $\phi-Higgs$ mixing angle. The right plot shows the reach of MATHUSLA, CODEX-b, FASER (analogous to Fig.~\ref{fig:sensitivity_SMS}), LHCb as well as direct detection experiments. LLP searches provide the only probe of large regions of DM parameter space. Figure from~\cite{DAgnolo:2018wcn}}
    \label{fig:coannihilation}
\end{figure}

Conversely to the ADM example, some DM models predict and rely on the existence of LLPs in a very instrumental fashion, but the LLP signature itself does not involve the DM particle directly. 
For example, the co-annihilation scenario studied by~\cite{DAgnolo:2018wcn} features a light scalar $\phi$ that mixes with the Higgs and provides the final state for $\chi_i \chi_j \to \phi \phi$ annihilations that set the abundance of the DM candidate $\chi_1$.
The properties of $\phi$ set the DM abundance, but the search for $\phi$ corresponds exactly to the search for the SM+S simplified model, see Fig.~\ref{fig:sensitivity_SMS}, and involves no DM production in itself. 
Even so, the LLP search can provide the most sensitive probe of the model, at far smaller couplings than is possible in direct detection experiments. This is demonstrated in Fig.~\ref{fig:coannihilation} from~\cite{DAgnolo:2018wcn}.

Another well-known example in this class is the $\nu$MSM~\cite{Asaka:2005pn, Asaka:2005an}. In this scenario, DM is a keV sterile neutrino produced in the early universe through the resonant Shi-Fuller mechanism~\cite{Shi:1998km}. The chemical potentials that are required for this resonance to occur are produced in the decays of the two heavier siblings of the DM sterile neutrino~\cite{Laine:2008pg}. Those heavier right-handed neutrinos are LLPs and can be searched for at MATHUSLA and other LLP experiments, see Fig.~\ref{fig:sensitivity_HNL}. MATHUSLA would therefore indirectly probe the mechanism of DM generation~\cite{Canetti:2012vf, Canetti:2012kh}.

Both of these last scenarios vividly demonstrate that simplified LLP scenarios like the HNL and SM+S model show up as components in more complete hidden sectors in general and dark matter models in particular, additionally motivating these searches. At the same time it is important to keep in mind that even in above scenarios like co-annihilation, other LLP signatures can easily be generated. 
The iDM scenario~\cite{Berlin:2018jbm} discussed above is one example, another is the related scenario of freeze-out through co-scattering~\cite{DAgnolo:2017dbv}, which can result in long-lived states that could only be produced in exotic Z decays~\cite{Aielli:2019ivi} at the LHC, where MATHUSLA could provide the best sensitivity.

Finally, there are many theories where DM is embedded in a larger hidden sector that, by its nature, is motivated to contain observable LLPs, even if  other aspects of the new physics determine the DM properties and abundance. 
For example, any composite hidden sector that contains a stable DM candidate will generically include a tower of slightly heavier states with possibly small mass splitting, meaning hidden sector production at colliders gives rise to LLP signals at collider-observable lifetimes~\cite{Li:2019ulz} with DM in the final states of the unstable particle decays. 
This is highly motivated in  SIMP and ELDER models~\cite{Hochberg:2014dra, Kuflik:2015isi}, where $3 \to 2$ annihilations set the relic abundance, and which are most easily realized in composite theories (though perturbative UV completions that give rise to LLPs also exist~\cite{Dery:2019jwf}).
Dynamical Dark Matter~\cite{Dienes:2011ja,Dienes:2011sa,Curtin:2018ees} is another example in this broad category, where a tower of unstable states of widely varying lifetimes make up the observed DM relic density, and shorter-lived states in the spectrum could be observable as LLPs at colliders.

The above discussion makes clear that LLP searches are a vital component of the DM search program. Depending on the DM scenario, it is possible that the properties or existence of LLPs directly determines the abundance of DM, or that the nature of the hidden sector which gives rise to DM also has to give rise to LLPs. 
In many cases, LLP decays to DM + SM are the only DM production mechanism that is available. 
This raises the question of how MATHUSLA could detect DM in LLP decays. We briefly discuss this below

%%%%%%%%%%%%%%%%%%%%%%%%%%%%%%%%%%
%%%%%%%%%%%%%%%%%%%%%%%%%%%%%%%%%%
%%%%%%%%%%%%%%%%%%%%%%%%%%%%%%%%%%
\subsection{Characterization of New Physics with MATHUSLA and CMS}
\label{s.LLPcharacterization}

\begin{figure}
    \centering
    \hspace*{-1mm}
     \includegraphics[width=0.8\textwidth]{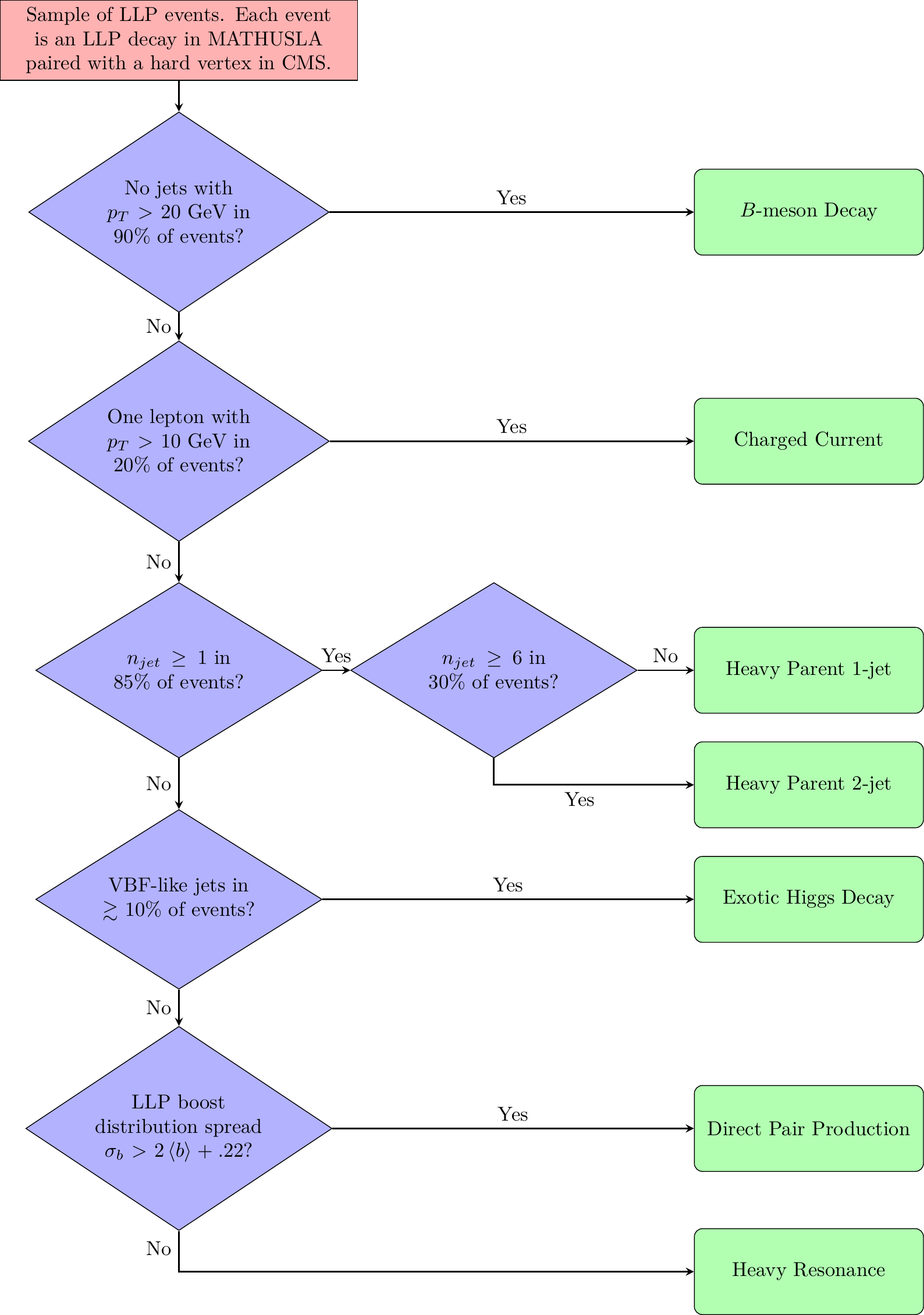}
     \caption{Summary of a prototype hierarchical LLP production mode classification algorithm. A sample of LLP observations is considered, and at each step a cut on a sample-level variable is used to classify the production mode. Extension to include additional LLP production modes, like parent paraticles decaying to LLP + leptons, or more general higgs-like scalars, is straightforward~\cite{Barron:2020kfo}.}
     \label{fig:prodmodeclassifier}
\end{figure}

Here we summarize several recent studies~\cite{Curtin:2017izq,Ibarra:2018xdl,Barron:2020kfo} that demonstrate MATHUSLA, working in tandem with the CMS main detector, can characterize the newly discovered physics in great detail, including determination of LLP decay mode, mass, production mode, and underlying parameters of the new physics theory like the parent particle mass. 
Unlocking this surprisingly broad analysis capability, given that MATHUSLA does not collect energy or momentum information, requires MATHUSLA to act as a Level-1 Trigger for CMS, allowing for the correlation of information from both detectors.

Let us start with the basic properties of the LLP in MATHUSLA. 
It has been shown~\cite{Curtin:2017izq} that
MATHUSLA can use geometric information from LLP decays to extract the likely velocity, or boost $b$, of the LLP, thereby determining the LHC bunch crossing that produced the LLP to be identified.
The multiplicity of final states in the DV strongly constrains the likely LLP decay modes, with leptonic/hadronic decays giving rise to low- or high-multiplicity final states. Refinements based purely on geometrical information as well as measurement of non-relativistic tracks are possible, including estimation of the hadronic decay's jet flavor~\cite{Curtin:2017izq} or the detection of an invisible component, possibly signifying the production of DM in an LLP decay~\cite{Ibarra:2018xdl}.

Determining the LLP mass, as well as other parameters of the underlying theory, is possible by correlating information from MATHUSLA with the main detector, and by observing that for a known LLP production mode, the measured boost of the LLP is highly correlated with its mass. 
This was recently studied in detail~\cite{Barron:2020kfo}, and we summarize the main results here.

If MATHUSLA can supply a L1 trigger signal to the main detector, each LLP decay reconstructed in MATHUSLA can be associated with an event recorded at CMS.\footnote{While uncertainty in the LLP velocity and pileup can both lead to a list of possible LLP production vertices, this is not prohibitive for the tasks of classifying the production mode and measuring model parameters~\cite{Barron:2020kfo}.}
Given a sample of observed LLPs, a simple classifier using cuts on the collective kinematic properties of the MATHUSLA-CMS events is sufficient to classify samples of LLP observations into one of the simplified LLP production models recently defined in~\cite{Alimena:2019zri}.
The production modes considered include 
exotic B meson decay to one LLP plus SM mesons and/or leptons (not included in the simplified models of~\cite{Alimena:2019zri} but added here as a stand-in for low-scale LLP production mechanisms),
``Exotic Higgs decay" to two LLPs,
a heavy vector-like resonance decaying to two LLPs (``Heavy Resonance"), 
a charged vector resonance decaying to one LLP and one SM charged lepton (``Charged Current"), 
a pair-produced heavy parent decay to an LLP plus SM jets (``Heavy Parent''), 
and ``Direct Pair Production'' of LLPs through an effective 4-point interaction.
The framework can also be easily included to include other production modes modes like heavy parent decay to leptons and LLPs or more general Higgs-like scalar decay to LLPs. 
Fig. \ref{fig:prodmodeclassifier} shows how the cuts are defined and sequentially applied in the classifier to discriminate between each production model. 
Assuming only 100 observed LLP events originating from a single production mode, samples can be accurately classified with probability from $90-100\%$, depending on the model, for a wide range of BSM particle masses and widths.

Once the LLP production topology is identified,  information from MATHUSLA and CMS can be used to determine the underlying parameters, most importantly LLP mass and, if applicable, the mass of the LLP parent particle. 
For Exotic $B$-decay, Exotic Higgs Decay and Direct Pair Production, the LLP boost distribution reveals the LLP mass at $\lsim 10\%$ precision with only 100 observed events. 
For the other production modes, an additional variable is needed in addition to LLP boost to determine both LLP mass and parent particle mass. 
A maximum likelihood fit in lepton $p_T$ (Charged Current) or jet $H_T$ (Heavy Parent) measured in the main detector together with the measured LLP boost distribution in MATHUSLA allows both parent and LLP masses to be determined with percent-level precision assuming only 100 observed events. 
The Heavy Resonance model is the most challenging due to the absence of additional visible objects in the main detector at lowest order. The LLP boost very accurately determines $m_{parent}/m_{LLP}$, but absolute determination of the parent mass scale requires either very high statistics ($\sim$1000 observed events) and exploiting the modest scale dependence of total jet energy in ISR, or direct observation of the intermediate resonance in visible resonance searches, which is expected to be available in the event of an LLP discovery. This is analogous to how the best bounds in simplified model of Dark Matter can come from resonance searches for the mediator~\cite{Alexander:2016aln}.

Overall, the recent analysis of~\cite{Barron:2020kfo}
together with studies of LLP decays at MATHUSLA alone~\cite{Curtin:2017izq,Ibarra:2018xdl}
demonstrates that MATHUSLA and CMS working together will be able to determine the properties of the LLP and the associated new physics in great detail, assuming MATHUSLA can act as a Level-1 trigger for CMS.

%%%%%%%%%%%%%%%%%
\section{Backgrounds to LLP Searches}
\label{s.backgrounds}
%%%%%%%%%%%%%%%%%

The new physics reach of MATHUSLA relies on its ability to distinguish LLP decays from the other particles that interact with or inside the detector. The vast majority of tracks observed in MATHUSLA will originate from cosmic rays and, at a much lower rate, energetic muons from the LHC. 
It is not obvious how any of these processes could fake an LLP decay, given the stringent geometric and timing requirements on the DV signal, but the high integrated rate of these backgrounds over the run of the HL-LHC necessitates a careful analysis. Atmospheric neutrinos can also scatter inelastically with atoms in the air-filled decay volume, a process which occurs at much lower rate than other backgrounds but which has the potential to be dangerous since it can give rise to a genuine DV. 
In this section, we provide a major update on the background estimates in~\cite{Chou:2016lxi, Alpigiani:2018fgd}. 
Data from the MATHUSLA Test Stand~\cite{Alidra:2020thg} and additional simulations provide a much clearer picture of the cosmic ray background in Section~\ref{s.cr}.
Detailed simulations of muon production at the LHC and propagation in the rock towards MATHUSLA for the updated geometry presented in this note are presented in Section~\ref{s.muon}.
Finally, Section~\ref{s.neutrinos} discusses full simulation studies of atmospheric neutrinos scattering off air in the MATHUSLA decay volume, updating earlier analytical calculations. 
These new results largely confirm earlier estimates that the background-free regime for LLP searches should be achievable, but also identify important areas of future detailed study that are currently being undertaken by the MATHUSLA collaboration.

\subsection{Cosmic Rays}
\label{s.cr}

The discussion of cosmic rays at MATHUSLA@CMS is calibrated by measurements taken  with the MATHUSLA Test Stand~\cite{Alidra:2020thg}.
The Test Stand had an active area of 2.5 m $\times$ 2.5 m and a vertical size of 6.5m, with three layers of RPCs and two layers of scintillator tiles providing tracking with timing information. 
As part of the analysis, cosmic rays incident on the detector were simulated with PARMA 4.0~\cite{PARMA3, PARMA4, PARMAsite}, 
together with a GEANT4 10.6~\cite{Agostinelli:2002hh} model of the detector and surrounding area to capture material interactions as the cosmic rays passed through the detector and hit the floor. 

\subsubsection{Downward Tracks}

The measured downward track rate was about 30 Hz~\cite{Alidra:2020thg}.
Simulations were used to estimate that the geometric acceptance of the Test Stand for downward traveling cosmic rays was $6\%$\footnote{A reconstructed track has to pass through all layers of the test stand detector, and its narrow vertical shape results in this low geometric acceptance.}
The track reconstruction efficiency was estimated from unbiased trigger events in data to be 40\%.
Taking this into account, the observed rate of downward tracks corresponds to a measurement of the local cosmic ray muon flux:
\begin{equation}
    \Phi_\mu^\mathrm{CR} \ \  \approx \ \ 1.2 \  \mathrm{cm}^{-2}  \ \mathrm{min}^{-1} \ ,
\end{equation}
in good agreement with expectation~\cite{Tanabashi:2018oca}.
This can be applied to the 100~m $\times$ 100~m area of MATHUSLA@CMS.
MATHUSLA has much greater geometrical acceptance than the test stand, and is also likely to have much better track reconstruction efficiency. We therefore assume that every cosmic ray hitting the top tracking layer of MATHUSLA can be reconstructed. This results in a downward CR rate of 2 MHz over the entire detector, or 
\begin{equation}
    N_\mathrm{down} \sim 3 \times 10^{14}
\end{equation}
CR tracks over the run-time of the HL-LHC, in agreement with earlier estimates. 
Downward-traveling CRs by themselves are extremely unlikely to fake an LLP: with a $\sim$ ns time resolution, imposing the requirement that a two-pronged DV is made up of only upwards going tracks rejects all cosmic rays at MATHUSLA, even before taking into account geometrical information~\cite{Chou:2016lxi, Alpigiani:2018fgd}. However, the CRs can give rise to other backgrounds that need closer examination.

\subsubsection{Upward Tracks}

CRs that hit the floor of the decay volume can produce upwards traveling SM particles through inelastic backscatter, or through the decay of stopped muons in the floor.
The measured ratio of upward tracks to downward tracks in the test stand without beam was 
\begin{equation}
\label{eq:NupoverNdown}
    \frac{N_\mathrm{up}^\mathrm{TS}}{N_\mathrm{down}^\mathrm{TS}} = (7.0 \pm 0.2) \cdot 10^{-5}.
\end{equation}
Different CR particle species produce upwards traveling tracks at different rates, necessitating detailed simulations to estimate the rate of upwards tracks.

Our simulations reproduced the measured ratio of upward tracks to downward tracks with very reasonable agreement. 
The composition of downward cosmic ray tracks was estimated in our simulations to be  49\% positive muons, 43\% negative muons, 4\% electrons, 3\% positrons, and 1\% protons.
The upward tracks induced by cosmic rays can be attributed to (by incident downward particle) 39\% neutrons, 20\% protons, 19\% positive muons, 10\% negative muons, 5\% positrons, 4\% photons, and 3\% electrons.
The upward tracks themselves are composed primarily of electrons, positrons, and charged pions.
Other particles such as protons, muons, and charged kaons collectively represent less than 5\% of the upward tracks.

There are two distinct populations of these upward tracks that can be distinguished by timing.
About 76\% of the tracks are consistent with the timing of an upward particle immediately created by the incident downward cosmic ray interacting with material in the test stand or the floor.
The remaining 24\% of tracks are delayed by tens to thousands of nanoseconds relative to the incident particle.
These are the result of low-energy muons stopping in or near the test stand and decaying, leading to a late upward electron or positron.

The measured ratio of upwards tracks to downward tracks in Eqn.~(\ref{eq:NupoverNdown}) can be readily applied to MATHUSLA@CMS, since geometrical acceptances and efficiencies largely drop out of the ratio.
This allows us to estimate that MATHUSLA would see roughly 
\begin{equation}
    N_\mathrm{up}\sim 2 \times 10^{10}
\end{equation}
upwards tracks over the HL-LHC run, mostly electrons, protons, and charged pions. 

This flux of upward tracks could fake a DV by random crossings. The rate of this background can be readily estimated, assuming the tracks occur randomly over the are of the detector and are uncorrelated.  Assuming two tracks have to pass within distance $\Delta d$ of each other in a time interval $\Delta t$ in order to pass  geometrical and timing DV reconstruction cuts, the number of ``fake'' DVs from random upward tracks is roughly
\begin{equation}
    N_\mathrm{fake\ DV} \sim  0.01 \times
    \left( \frac{N_{up}}{10^{10}} \right)^2
    \left( \frac{\Delta d}{10 \mathrm{cm}} \right)^2 
    \left( \frac{\Delta t}{10 \mathrm{ns}} \right)
\end{equation}
Given that MATHUSLA's trackers will have $\lsim$ 1-10 cm (depending on the orientation of the tracking plane) and ns spatial and timing resolution, this suggests that random crossings will not be a dominant source of background for LLP searches, though suppressing this background may become one of the drivers of tracking resolution as the design is further developed.

Other possible sources of backgrounds depend on the identity of the upwards traveling SM particles themselves. 
For example, charged pions make up a large fraction of upward traveling particles and have a rare decay, $\mathrm{Br}(\pi^+ \to e^+ e^- e^+ \nu_e) \approx 3 \times 10^{-9}$~\cite{Tanabashi:2018oca}, that could fake a multi-pronged DV from BSM LLP decays $\mathcal{O}(1)$ times throughout the entire MATHUSLA run.
Muons, which constitute a much smaller fraction of upwards traveling particles, also have a rare decay 
$\mathrm{Br}(\mu^- \to e^- e^+ e^- \nu_\mu \bar \nu_e) \approx 3 \times 10^{-5}$
that could contribute to LLP backgrounds at a similar rate. 
Finally, neutral kaons make up a tiny but as yet undetermined fraction of produced particles. They have a dominant decay to three charged particles, making them another possible LLP background. 

The above discussion must be regarded as strictly preliminary. 
Reliably estimating the rates of these extremely rare processes requires careful study and  high-statistics simulations to estimate the muon and especially $K_0^L$ production rates from cosmic rays incident on the detector floor. 
This can then be combined with the known decays of these particles to arrive at a more complete picture of these ultra-rare backgrounds to LLP searches. Several veto-strategies are available, which depend on the spectrum of the produced upwards-traveling particle species as well as the precise design of the floor detector, which could be optimized to help reject these backgrounds by detecting the originating cosmic ray hit. The MATHUSLA collaboration is currently in the process of conducting the required studies.

\subsection{Muons from HL-LHC Collisions}
\label{s.muon}

In this section, we provide an estimate for the rate of upwards traveling Muons produced at the HL-LHC and detected by MATHUSLA.  
Compared to earlier estimates, these simulations use a more accurate material description of the rock and CMS cavern.  Furthermore, we now only count muons that pass through scintillator layers, which gives us a better estimate of the geometric acceptance.
These simulations, run with the appropriate geometrical differences, closely reproduce the number of upwards tracks  observed in the MATHUSLA test stand while with the beam on~\cite{Alidra:2020thg}.

High energy muons with enough energy to reach MATHUSLA are dominantly produced in $W$-production and $\bar b b$ production at the HL-LHC, which is simulated in PYTHIA8~\cite{Sjostrand:2007gs}.
The $b\bar{b}$ production cross-section is obtained directly from PYTHIA8, while the W production cross-section at 14 TeV is extrapolated from the measured 13 TeV $W$-production cross-section~\cite{Aad:2016naf} using MadGraph~\cite{Alwall:2011uj} to extract the LO dependence on $\sqrt{s}$. 
The new GEANT4 simulation~\cite{Agostinelli:2002hh} propagates muons through three distinct layers of rock from the IP to the MATHUSLA detector, and includes details of the experimental cavern UXC 55,   the PX56 shaft, and the modular MATHUSLA@CMS detector design. 

The number of muons observed by MATHUSLA is defined as the number of events with Muons that hit the detector's top seven layers. 
Assuming an instantaneous luminosity of $10^{35}$ $cm^{-2} s^{-1}$ at the HL-LHC, we obtain  a rate of about $1.03 \times 10^4$ ($1.02\times 10^4$) muons per hour from $W$-production ($\bar b b$ production with $p_T^b > 40 \gev$). 
Scaling up the total number slightly to account for small contributions from $Z$-bosons, tau- and charm-decays, we obtain a total rate of approximately 
\begin{equation}
    N_{\mu} \approx \left( \frac{2.4 \times 10^{4}}{\mathrm{hour}} \right) \ \times \ 
    \left(
    \frac{\mathcal{L}}{
    10^{35} \  cm^{-2} \ s^{-1}
    }
    \right)
\end{equation}
at MATHUSLA. The total number of muons from HL-LHC collisions passing through MATHUSLA @ CMS, assuming integrated luminosity of 3000fb$^{-1}$, is therefore 
\begin{equation}
    N_\mu \approx 2 \times 10^8 \ .
\end{equation}
This is about 50~$\times$ more than the rate estimated in earlier background estimates for the MATHUSLA200 geometry~\cite{Alpigiani:2018fgd}, which has similar solid angle coverage (consistent with the near-identical acceptance for LLP decays). 
This can be understood as arising from the smaller amount of rock separating MATHUSLA @ CMS from the LHC collision compared to MATHUSLA200. 
It happens to be the case that a muon must have kinetic energy close to $m_W/2$ to penetrate about 100m of rock. 
For MATHUSLA200, this means that $W$ decay dominates the upward muon rate. Reducing the amount of material drastically increases the muon contribution from $W$ and $Z$ decay due to the kinematic edge in the distribution of muon momenta. It also allows softer muons from $\bar b b$ production to contribute in comparable numbers.

Upwards LHC muons do not constitute a background to LLP searches. 
However, the muon background studies in the LOI~\cite{Alpigiani:2018fgd} include predictions for rare events arising from the muons flying through the decay volume, like highly subdominant rare muon decay modes. %
Rescaling those rates by 50 suggests that MATHUSLA @ CMS might see $\mathcal{O}(10)$ $\mu \to e e e \nu \nu$ decays, which would have to be vetoed using the floor detector or main detector information. 
Material interactions can be rejected using a fiducial veto. 
Future studies will refine on these estimates and explicitly demonstrate the associated rejection strategies.

\subsection{Neutrinos}
\label{s.neutrinos}

Neutrinos can scatter off nuclei in the air-filled MATHUSLA decay volume, giving rise to charged particles that go through the tracking layers and mimic an LLP decay inside the detector. This background category can be further divided into neutrinos from the LHC, and those from cosmic ray interactions in the atmosphere. Previous calculations \cite{Chou:2016lxi} estimated that the rate of these background processes is very low. We have now conducted detailed calculations using GENIE \cite{Andreopoulos:2009rq} to simulate neutrino interactions confirm this to be the case.

Neutrinos were simulated using GENIE v2.12.10 to generate neutrino events in the energy range 0-10 GeV incident on an air target (76.8\% N, 23.2\%O), using GENIE's pre-calculated DefaultPlusMECWithNC set of interaction cross-sections. Samples were generated for muon and electron neutrinos and anti-neutrinos, and reweighted according to the total neutrino flux $d N_\nu/dE_\nu$ incident on the MATHUSLA detector, as we discuss below.

The atmospheric neutrino background was estimated using flux measurements from the Fréjus experiment \cite{Frejus:1995flux}. 
Based on the simulations, the expected number of neutrino interactions occurring within the volume of MATHUSLA100 over one year is $\approx$ 30, assuming the detector runs for 50\% of the year. A fairly modest cut asking for each event to have two tracks with $\beta >$ 0.8 reduces this number to $<$ 1 expected background event from atmospheric neutrinos per year. 
The current estimation does not include any geometrical cuts, and the background can be further reduced by requiring that the vertex formed by these tracks points back to the IP. Such cuts would depend both on analysis strategies and the resolution of the detector.

The other category of neutrino backgrounds is from neutrinos that are produced at the LHC and scatter off nuclei within the MATHUSLA detector volume. We estimate this background by simulating minimum-bias collisions at the LHC and calculating the number of interactions expected inside MATHUSLA100 from resulting neutrinos. 
The rate of this background is extremely small. With no cuts and assuming an integrated luminosity of 3000 fb$^{-1}$ over the complete HL-LHC run, the total number of neutrino events expected is $<$ 0.52. Asking for two tracks with $\beta >$ 0.8 further reduces this to $\approx$ 0.01 events.
Neutrinos therefore do not constitute a significant background to LLP searches at MATHUSLA.

%%%%%%%%%%%%%%%%%
\section{Cosmic Ray Science Program}
\label{s.cosmicrays}
%%%%%%%%%%%%%%%%%

MATHUSLA's large detection area and highly granular tracking system makes it an excellent cosmic-ray (CR) telescope to study extended air showers (EAS) and performing CR measurements up to the PeV scale. 
Highly detailed measurements of EAS arrival time, particle multiplicity, and spatial distributions allow reconstruction of EAS core and direction and would be greatly aided by installing a hybrid digital-analogue tracking layer of Resistive Plate Chambers (RPC) similar to the ARGO-YBJ experiment. 
Such a possible upgrade is now being closely investigated, since CR measurements constitute a guaranteed physics return on the investment of building MATHUSLA. 
This will be discussed in an upcoming MATHUSLA publication~\cite{MathuslaCRWP}, and is an area where new members with cosmic ray expertise could make significant contributions.

%%%%%%%%%%%%%
\section{Next Steps}
\label{s.nextsteps}

The MATHUSLA collaboration plans to finalize a Technical Design Report for the design of MATHUSLA, including a robust cost estimate, in 2021. 
The plan for the following years includes the construction of a prototype detector module that can already set relevant limits on certain LLP scenarios and conduct cosmic ray measurements during the extended LHC shutdown to further calibrate background predictions. 
Subsequent construction and commissioning of the remaining detector modules should then allow the full detector to come online in time for first beam at the HL-LHC. 

To achieve these goals, significant R\&D is needed to design and optimize the detector hardware itself, including the scintillators, wavelength shifting fibers, SIPMs, DAQ and trigger systems. 
Detailed simulation studies including full material effects are needed to demonstrate LLP decay reconstruction and explicit background rejection in the unique environment of MATHUSLA.
Similarly, it is a high priority to conduct careful analyses of rare backgrounds like production of upward traveling muons, pions and particularly neutral kaons, as well as studying their associated veto strategies and possible implications for the detector design and geometry. 
Finally, studies are underway to understand the cosmic ray physics case, and the implications of a possible hybrid digital-analog detection layer to reconstruct dense air shower cores.

The collaboration is accepting new members, and new contributions are particularly welcome in all of the above areas, from detector hardware to simulation and cosmic ray physics.

%%%%%%%%%%%%%%%%%
\bibliography{references}
\bibliographystyle{JHEP}
%%%%%%%%%%%%%%%%%

\end{document}